\documentclass[sigconf, authorversion=true]{acmart}
\pdfoutput=1

\usepackage{booktabs} %

\usepackage{listings,multicol}

\lstdefinestyle{2col}{tabsize=2,language=python,basicstyle=\footnotesize,commentstyle=\scriptsize\ttfamily,numbers=left,numberstyle=\tiny,firstnumber=last,mathescape,morekeywords={struct,integer,boolean},numberblanklines=false,multicols=2,xleftmargin=0.6cm,escapechar=~,columns=flexible
}

\lstdefinestyle{1col}{tabsize=2,language=python,basicstyle=\footnotesize,commentstyle=\scriptsize\ttfamily,numbers=left,numberstyle=\tiny,firstnumber=last,mathescape,morekeywords={struct,integer,boolean},numberblanklines=false,xleftmargin=.5cm,escapechar=~,
columns=fullflexible}

\usepackage{tikz}
\usetikzlibrary{%
decorations.pathreplacing,%
decorations.pathmorphing,%
decorations.shapes,%
decorations.text,%
decorations.markings,%
shapes,%
shapes.callouts,%
shadows,%
arrows,
calc,%
positioning,%
chains,%
backgrounds,%
fit, %
fadings, %
shapes.multipart}

\usepackage[export]{adjustbox}
\usepackage[caption=false,font=footnotesize]{subfig}

\usepackage{algorithm}

\copyrightyear{2018}
\acmYear{2018}
\setcopyright{acmcopyright}
\acmConference[SPAA '18]{30th ACM Symposium on Parallelism in Algorithms and Architectures}{July 16--18, 2018}{Vienna, Austria}
\acmBooktitle{SPAA '18: 30th ACM Symposium on Parallelism in Algorithms and Architectures, July 16--18, 2018, Vienna, Austria}
\acmPrice{15.00}
\acmDOI{10.1145/3210377.3210408}
\acmISBN{978-1-4503-5799-9/18/07}

\newcommand{\PSIM}{{\sf PSim}} %
\newcommand{\LFSplit}{{\sf LF-Split}} %
\newcommand{\LFSplitU}{{\sf LF-Split-U}} %
\newcommand{\LFSplitM}{{\sf LF-Split-M}} %
\newcommand{\WFFreeze}{{\sf WF-Freeze}} %
\newcommand{\LFFreeze}{{\sf LF-Freeze}} %
\newcommand{\LFFreezeU}{{\sf LF-Freeze-U}} %
\newcommand{\LFFreezeM}{{\sf LF-Freeze-M}} %
\newcommand{\WFExt}{{\sf WF-Ext}} %
\newcommand{\Lock}{{\sf Lock}} %

\newcommand{\CAS}{{\tt CAS}}
\newcommand{\DCAS}{{\tt DCAS}}

\newcommand{\ADD}{{\tt Add}}

\newcommand{\ignore}[1]{}

\newcommand{\remove}[1]{}

\newcommand{\LOOKUP}{{\sc LookUp}}
\newcommand{\INSERT}{{\sc Insert}}
\newcommand{\DELETE}{{\sc Delete}}

\begin{document}

\title{An Efficient Wait-free Resizable Hash Table}

\author{Panagiota Fatourou}
\affiliation{%
  \institution{FORTH ICS \& University of Crete, Department of Computer Science}
  \country{Greece}
}
\email{faturu@csd.uoc.gr}
\author{Nikolaos D. Kallimanis}
\affiliation{%
  \institution{FORTH ICS}
  \country{Greece}
}
\email{nkallima@ics.forth.gr}
\author{Thomas Ropars}
\affiliation{%
  \institution{Univ. Grenoble Alpes}
  \country{France}
}
\email{thomas.ropars@univ-grenoble-alpes.fr}

\begin{abstract}

  This paper presents an efficient wait-free resizable hash table. To
  achieve high throughput at large core counts, our algorithm is
  specifically designed to retain the natural parallelism of
  concurrent hashing, while providing wait-free resizing. An extensive
  evaluation of our hash table shows that in the common case where
  resizing actions are rare, our implementation outperforms all
  existing lock-free hash table implementations while providing a
  stronger progress guarantee.

\end{abstract}

\date{}

\maketitle

\section{Introduction}

As the core count is increasing in modern processors, designing data
structures that provide good performance when a large number of
threads access them concurrently is a must~\cite{Shavit2011}, but it
is also a highly challenging task~\cite{Gramoli2015}, especially if
strong liveness guarantees, such as {\em wait-freedom}, are to be
provided~\cite{Herlihy1991}. Wait-freedom is highly desirable as it
ensures \emph{progress} for all running threads independently of their
speeds or any {\em crash} failures that other threads may experience.
Specifically, wait-freedom ensures that {\em every} operation on the
data structure (executed by a running thread) will complete within a
finite number of steps.  \emph{Lock-freedom} is a weaker progress
condition which guarantees that {\em at least one} thread makes
progress, thus allowing other threads to starve.

A hash table is a data structure commonly used to implement a
dictionary of key-value pairs. It provides two \emph{update}
operations (\INSERT\ and \DELETE ) and a \LOOKUP\ operation.  A hash function is
used to associate keys to buckets so that each operation on
the hash table takes constant average time~\cite{Herlihy2008}.  To
ensure this property even when the number of stored items varies over
time, dynamic hashing aims at dynamically resizing the hash table to
adapt the number of buckets to the number of
items~\cite{Enbody1988}. Resizing actions (splitting or merging of
buckets) are triggered during \INSERT\ and \DELETE\ operations.

It is commonly acknowledged that, in most cases, \LOOKUP s are by far
the most frequent operations on a hash table~\cite{Herlihy2008,
  Triplett2011}. Given that update operations are not frequent,
resizing actions are rare events since the number of items to store does
not vary much over time. In theory, hashing can be made very efficient
in a concurrent environment due to its \emph{natural
  parallelism}~\cite{Herlihy2008}. In most cases, concurrent
operations access different parts of the hash table, and thus
they can proceed in parallel without any interference with one
another.  However, the case of dynamic hashing is more complex.
Allowing hash table operations to be executed in parallel with
resizing actions while ensuring {\em
  linearizability}~\cite{Herlihy1990}, is difficult, especially for
algorithms providing strong progress guarantees~\cite{shalev2006,
  Liu2014}.

This paper presents a new wait-free implementation of a resizable hash
table. Our hash table aims at achieving best performance for the most
common operations on a hash table, while providing {\em wait-freedom}
progress guarantee. To guide our design, we identify two design rules
that are important to achieve high performance at large core counts:
(A) \LOOKUP\ operations should always be allowed to proceed without
any synchronization; (B) when no resizing actions are executed, update
operations applying to different buckets should be allowed to progress
fully in parallel (i.e., without any interference with each
other). These rules aim at preserving the natural parallelism of hash
tables in the most frequent case where no resizing is required. Rule
(A) additionally aims at minimizing the cost of the most frequent
operations in all cases.

To implement a hash table that complies with these design rules, we
propose an algorithm based on {\em extendible hashing}, a dynamic
hashing technique that considers keys as bit
strings~\cite{Fagin1979}. An extendible hash table can be seen as an
array (the {\em directory}) of pointer to fixed-size buckets. In its
sequential implementation, every resizing operation is local, e.g.,
one bucket can be split into two without modifying the other buckets.

Our wait-free extendible hash table relies on the \PSIM\ universal
construction~\cite{Fatourou2013}. \PSIM\ provides a
general mechanism to implement any concurrent object
in a wait-free manner. 
It exploits the well-known technique~\cite{Fatourou2012} of
having a thread that executes an operation help other 
announced operations 
by applying them, in addition to its own,
on a local copy of the simulated object state. 
Then, it attempts to change a shared reference to the object state
 to point to this local copy. 
\PSIM\ results in 
highly-efficient wait-free implementations of 
data structures that have a single or a small number
of points of contention, such as stacks and
queues~\cite{Fatourou2013}.
In this paper, we show how we can use
several instances of the \PSIM\ algorithm and synchronize them
appropriately to get an efficient wait-free implementation of an
extendible hash table.  To match our design rules, our hash table uses
an instance of \PSIM's algorithm for each bucket. These instances run
update operations on each bucket fully independently as long as no
resizing actions are required. An additional instance of \PSIM's
algorithm is used to manage resizing actions modifying the state of
the hash table.  The crux of our algorithm is in the mechanisms used
to coordinate the different instances of \PSIM\ during
resize actions in order to ensure linearizability and wait-freedom,
while complying with our design rules.

We evaluate our algorithm experimentally and compare it with that of
two state-of-the-art lock-free concurrent hash tables: the lock-free
resizable hash table proposed by Liu et al.~\cite{Liu2014}
and the hash table based on a \emph{split-ordered list} proposed by
Shalev and Shavit~\cite{shalev2006}. Experiments run on two Intel
processors (48-core Haswell, and 64-core Broadwell) show that in a
directory-stable state, i.e., when resizing actions are rare,
the performance of our algorithm is highly competitive. When
manipulating a small number of items, and when the percentage of
\LOOKUP s is high, our wait-free algorithm outperforms the most
efficient lock-free algorithm by up to 47\% on a 64-core machine. When
the size of the hash table increases, a modified version of the
\emph{lock-free} algorithm presented by Liu et
  al.~\cite{Liu2014}, that we propose, becomes the most efficient
with our algorithm being the second most efficient. The very high
performance of our algorithm in directory-stable states comes at the
cost of slower resizing actions. However, our experiments demonstrate
that the resizing cost is acceptable when it can be amortized over
long runs. In this case, our wait-free hash table is as efficient as
or more efficient than the best existing lock-free solutions.

The main contributions of this paper are summarized as follows:
\begin{itemize}
\item We present a new wait-free resizable hash table based on
  extendible hashing (Section~\ref{sec:algo}). Its implementation
  follows two design rules that aim at preserving the natural
  parallelism of dynamic hashing for the most frequent operations.
\item We provide an extensive performance evaluation of the new
  algorithm at large core counts that shows that, when resizing
  actions are rare, our wait-free hash table largely outperforms all
  previous lock-free implementations (in terms of throughput), at the
  cost of slower resizing (Section~\ref{sec:eval}).

\end{itemize}

The results of our experiments show evidence that the design rules 
we identify are of key importance to build efficient non-blocking
resizable hash tables.

\section{Related Work}

Hash tables are important data structures in domains ranging from
operating systems' kernels~\cite{Boyd-Wickizer2010, Triplett2011} to
runtime and programming languages~\cite{Jenkins2017}.

The easiest way to implement a resizable hash table is probably using
locks, i.e., in a {\em blocking} way.  One such
implementation is the \emph{ConcurrentHashMap} provided in the Java
{\tt concurrency.utils} library~\cite{Lea2003}. It uses a fixed number
of locks, each of which guards a subset of the buckets.  Resizing can
be performed only by a thread that has acquired all locks, thus
excluding all other threads from executing operations during a
resizing phase. A concurrent \emph{extendible} hashing algorithm based
on locks was presented by Ellis~\cite{Ellis1983}. It is
based on a two-level locking scheme where a lock on the directory must
be grabbed first, before locking a specific bucket.

Non-blocking resizable hash tables are appealing because of their
stronger progress guarantees that have been shown to lead to higher
performance in several studies~\cite{Gramoli2015,shalev2006}. One of
the first lock-free non-resizable hash table implementations was
described by Michael~\cite{Michael2002}. It is based on an array of
lock-free linked lists. At the same period,
Greenwald~\cite{Greenwald2002} presented a lock-free resizable hash
table that relied on a \DCAS\ (\emph{double-compare-and-swap})
operation. However, \DCAS\ is not supported on most hardware
architectures.

The lock-free resizable hash table proposed in \cite{shalev2006},
which we will call \LFSplit, relies on a \emph{split-ordered}
list. The items of the hash table are stored in an ordered linked-list
and a separate array of pointers pointing to elements in the linked
list plays the role of the directory. The items belonging to bucket
numbered $i$ are all the items accessible in the linked-list starting
from the node pointed to by entry $i$ of the directory and finishing
at the node pointed to by the next entry of the directory. Hence,
inserting a new element in a bucket simply requires to insert it in
the linked list. Splitting a bucket into two only implies adding in
the directory a new pointer to an element in the list. Relying on a
list to implement the hash table makes resizing operations very
efficient. However, \LOOKUP\ operations might be less efficient than
with array-based hash tables because of the cost of pointer chasing
paid when iterating over the elements of a bucket~\cite{Liu2014,
  David2015}. \LFSplit\ complies with neither of the design rules we
introduce to optimize directory-stable-state performance. Indeed, rule
(A) is violated because when a node is marked for deletion, threads
running \LOOKUP\ operations might have to help removing these nodes
from the list. Moreover, a global counter has to be updated when an
item is removed or a {\em new} item is inserted, which breaks rule
(B). The experiments presented in Section \ref{sec:eval} show that
these two facts limit the performance of \LFSplit\ when the directory
is stable.

To the best of our knowledge, only two wait-free resizable hash tables
have been described thus far~\cite{Feldman2013, Liu2014}. The
algorithm of Feldman et al.~\cite{Feldman2013} is based on a
multi-level array resulting in pointer chasing and therefore reduced
performance for \LOOKUP\ operations.

Liu et al. have proposed in~\cite{Liu2014} an array-based
resizable lock-free hash table, which we will call \LFFreeze. Buckets
are implemented as arrays of items, and the directory is an array of
pointers to items. When the directory needs to be resized, the buckets
are \emph{frozen}: no update operations are allowed to proceed on
these buckets anymore. Splitting the buckets is then lazily done
during \INSERT\ operations. \LFFreeze\ respects the two design rules
that we have defined.  The experimental analysis on large multicore
machines that is provided in~\cite{Liu2014}, shows that
\LFFreeze\ outperforms \LFSplit~\cite{shalev2006}. Our algorithm
shares some of the design ideas of \LFFreeze. However, it ensures
wait-freedom by using instances of the \PSIM\ algorithm for ensuring
synchronization.

Liu et al. also proposed in~\cite{Liu2014} a wait-free
variant of their lock-free implementation, which we will call
\WFFreeze.  In this variant, threads should help each other running
update operations and resize actions on the buckets. This is done by
assigning sequence numbers to all update operations, a technique that
is implemented using a globally shared counter (thus violating rule
(B)). The experiments in~\cite{Liu2014} show that
\WFFreeze\ exhibits performance that is by far lower than the
performance of \LFFreeze. Our algorithm ensures wait-freedom without
using any globally shared counter.

Kogan and Petrank proposed in~\cite{Kogan2012} the
\emph{fast-path-slow-path} technique, an approach to get a wait-free
variant of a lock-free algorithm with a relatively small performance
cost. In their experimental analysis, Liu et al.  included a
wait-free variant of their lock-free algorithm based on this
technique. The resulted algorithm performs much better than
\WFFreeze\ (their brute-force wait-free solution), but its performance
is lower than that of \LFFreeze\ (their lock-free algorithm).  Our
evaluation shows that our brute-force wait-free implementation
outperforms \LFFreeze\ in several cases when the hash table is in a
directory-stable state.

\section{Extendible hashing \label{sec:extendible}}

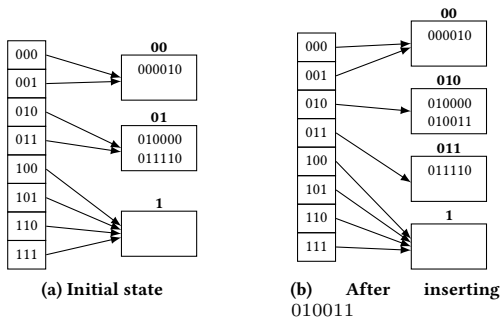
\begin{figure}[t]
  \begin{center}
    \subfloat[Initial state]{
      \label{fig:hash_init}
      \begin{tikzpicture}[
font=\scriptsize,
node distance=0,
text height=1ex, text depth=.25ex,
start chain=1 going below, node distance=-0.15mm]

  \node(e0) at (0,0) [draw, on chain=1] {000};
  \node(e1) [draw, on chain=1] {001};
  \node(e2) [draw, on chain=1] {010};
  \node(e3) [draw, on chain=1] {011};
  \node(e4) [draw, on chain=1] {100};
  \node(e5) [draw, on chain=1] {101};
  \node(e6) [draw, on chain=1] {110};
  \node(e7) [draw, on chain=1] {111};

  \coordinate (B01) at ($(e3.east)+(1,0)$);
  \coordinate (B00) at ($(e0.east)+(1,-0.2)$);
  \coordinate (B1) at ($(e6.east)+(1,0)$);
  
  \node(b01) at (B01) [rectangle, anchor=west, text
    width=.8cm, align=center, minimum height=15] {010000 011110};

  \draw ($(B01)+(0,0.2)$) -- node[midway, above=-0.1cm]
        {\textbf{01}}+(1,0) -- +(1,-0.6) -- +(0,-0.6) --
        cycle;

  \node(b00) at (B00) [rectangle, anchor=west, text
    width=.8cm, align=center, minimum height=15] {000010};

  \draw ($(B00)+(0,0.2)$) -- node[midway, above=-0.1cm]
        {\textbf{00}}+(1,0) -- +(1,-0.6) -- +(0,-0.6) --
        cycle;

  \node(b1) at (B1) [rectangle, anchor=west, text
    width=.8cm, align=center, minimum height=15] {};
  
  \draw ($(B1)+(0,0.2)$) -- node[midway, above=-0.1cm]
        {\textbf{1}}+(1,0) -- +(1,-0.6) -- +(0,-0.6) --
        cycle;

  \draw[-latex] (e4.east) -- ($(B1)+(0,0.0)$);
  \draw[-latex] (e5.east) -- ($(B1)+(0,-0.05)$);
  \draw[-latex] (e6.east) -- ($(B1)+(0,-0.1)$);
  \draw[-latex] (e7.east) -- ($(B1)+(0,-0.15)$);

  \draw[-latex] (e2.east) -- ($(B01)+(0,-0.1)$);
  \draw[-latex] (e3.east) -- ($(B01)+(0,-0.15)$);

  \draw[-latex] (e0.east) -- ($(B00)+(0,-0.1)$);
  \draw[-latex] (e1.east) -- ($(B00)+(0,-0.15)$);

\end{tikzpicture}
    }
    \hspace{1cm}
    \subfloat[After inserting $010011$]{
      \label{fig:hash_after}
      \begin{tikzpicture}[
font=\scriptsize,
node distance=0,
text height=1ex, text depth=.25ex,
start chain=1 going below, node distance=-0.15mm]

  \node(e0) at (0,0) [draw, on chain=1] {000};
  \node(e1) [draw, on chain=1] {001};
  \node(e2) [draw, on chain=1] {010};
  \node(e3) [draw, on chain=1] {011};
  \node(e4) [draw, on chain=1] {100};
  \node(e5) [draw, on chain=1] {101};
  \node(e6) [draw, on chain=1] {110};
  \node(e7) [draw, on chain=1] {111};

  \coordinate (B00) at ($(e0.east)+(1,0.15)$);
  \coordinate (B010) at ($(B00)+(0,-0.9)$);
  \coordinate (B011) at ($(B00)+(0,-1.8)$);
  \coordinate (B1) at ($(B00)+(0,-2.7)$);

  \node(b00) at (B00) [rectangle, anchor=west, text
    width=.8cm, align=center, minimum height=15] {000010};
  
  \draw ($(B00)+(0,0.2)$) -- node[midway, above=-0.1cm]
        {\textbf{00}}+(1,0) -- +(1,-0.6) -- +(0,-0.6) --
        cycle;

  \node(b1) at (B1) [rectangle, anchor=west, text
    width=.8cm, align=center, minimum height=15] {};
  
  \draw ($(B1)+(0,0.2)$) -- node[midway, above=-0.1cm]
        {\textbf{1}}+(1,0) -- +(1,-0.6) -- +(0,-0.6) --
        cycle;

  \node(b010) at (B010) [rectangle, anchor=west, text
    width=.8cm, align=center, minimum height=15] {010000 010011};

  \draw ($(B010)+(0,0.2)$) -- node[midway, above=-0.1cm]
        {\textbf{010}}+(1,0) -- +(1,-0.6) -- +(0,-0.6) --
        cycle;
        
  \node(b011) at (B011) [rectangle, anchor=west, text
    width=.8cm, align=center, minimum height=15] {011110};

  \draw ($(B011)+(0,0.2)$) -- node[midway, above=-0.1cm]
        {\textbf{011}}+(1,0) -- +(1,-0.6) -- +(0,-0.6) --
        cycle;

  \draw[-latex] (e0.east) -- ($(B00)+(0,-0.1)$);
  \draw[-latex] (e1.east) -- ($(B00)+(0,-0.15)$);
      
  \draw[-latex] (e4.east) -- ($(B1)+(0,0.0)$);
  \draw[-latex] (e5.east) -- ($(B1)+(0,-0.05)$);
  \draw[-latex] (e6.east) -- ($(B1)+(0,-0.1)$);
  \draw[-latex] (e7.east) -- ($(B1)+(0,-0.15)$);

  \draw[-latex] (e2.east) -- ($(B010)+(0,-0.1)$);

  \draw[-latex] (e3.east) -- ($(B011)+(0,-0.15)$);

\end{tikzpicture}
    }
    \caption{An extendible hash table\label{fig:hash}. The directory
      has a depth of 3 and each bucket can store at most 2 items.}
  \end{center}
\end{figure}

An {\em extendible hash table} is a data structure of two levels.  It
consists of a set of {\em buckets}, in each of which a fixed number
$b$ of items are stored, and a resizable array, called the {\em
  directory}, where each entry stores a pointer to a bucket.
Extendible hashing manipulates hash keys as bit strings: the bit
strings are used to distribute items to buckets.
The directory has $2^D$ entries, where $D$ is a parameter called 
the {\em depth} of the directory.
Figure~\ref{fig:hash} describes an extendible hash table
with $D = 3$. A prefix corresponding to the $D$ most significant
bits of the hash key is used to associate items to directory entries
and the buckets they point to.  For instance, in Figure~\ref{fig:hash}, key
$010000$ is associated with the directory entry $010$. 
The depth of the directory is at most as big as the total number of bits 
in the hash key. The size of the directory is always exponential to its depth.

In practice, the number of buckets does not have to be the same as
the number of directory entries: to improve memory efficiency, it is
adapted to the number of items to store.  In
Figure~\ref{fig:hash_init}, a single bucket is allocated to store all
keys with prefix $1$ since there is no such key, and all four
directory entries with prefix $1$ point to that bucket. The {\em
  depth} of this bucket is $1$, i.e., it equals the length of the
prefix that identifies the hash keys that are to be stored in it. 

As illustrated in Figure~\ref{fig:hash_after}, a bucket is split in
two buckets when it is full (i.e., when it already contains $b$ items)
and a newly inserted item must be stored in this bucket.  For
instance, to insert key $010110$, two new buckets $010$ and
$011$ should be created to replace the existing bucket $01$. 
In this case, the directory will not be resized 
because the bucket with prefix $01$ has depth $2$ and the two newly
created buckets to replace it, have depth $3$ which is smaller
than or equal to the directory depth.  However, to insert
another item $010111$ in the table of Figure~\ref{fig:hash_after},
resizing the directory would be required. The depth of the directory
should be increased to $4$ to allow storing the pointers of two new
buckets $0100$ and $0101$.
We remark that the resizing actions are \emph{local}: to
replace a bucket with two new buckets, all the items of the old
bucket are stored in the new buckets and no additional elements
are stored in them.

\section{A wait-free implementation of an extendible hash table} 
\label{sec:algo}

This section presents the new wait-free implementation of the
  extendible hash table. We start by describing the main ideas of the
  algorithm. Then, we present how the hash table works when no resizing
  occurs. 
  Finally, we provide the details of resizing.
Our wait-free algorithm incorporates and builds upon the code of \PSIM~\cite{Fatourou2013}
for updating the buckets and the directory. 

In our implementation (as well as in the experimental analysis
of Section~\ref{sec:eval}), an invocation of \INSERT\ for an
already existing key, updates the value associated with the key.  This
semantics correspond to that of dictionaries provided by popular
programming languages such as Java\footnote{Package
  \texttt{java.util.AbstractMap}} or
Python\footnote{\url{https://docs.python.org/3/library/stdtypes.html\#dict}}.
The description assumes that a wait-free garbage
collector is available.  (Section~\ref{sec:implem} provides a
discussion on memory reclamation.)  We consider a system of $n$
threads.

\subsection{The algorithm in a nutshell\label{sub:nutshell}}

\begin{figure}[t]
  \begin{center}
    \subfloat[Initial state\label{fig:table_init}]{
      \begin{tikzpicture}[
font=\scriptsize,
node distance=0,
text height=1ex, text depth=.25ex,
start chain=1 going below, node distance=-0.15mm]

  \def\distA{0.7}
  \def\distB{0.5}
  \def\distC{0.25}
  \def\sizeD{0.8cm}
  
  \node(e0) at (0,0) [draw, on chain=1] {00};
  \node(e1) [draw, on chain=1] {01};
  \node(e2) [draw, on chain=1] {10};
  \node(e3) [draw, on chain=1] {11};

  \node(dstate) at ($(e3.south)+(0,-0.2)$) {\texttt{DState}};

  \coordinate (B0) at ($(e0.east)+(\distA,-0.2)$);
  \coordinate (B1) at ($(e2.east)+(\distA,-0.2)$);

  \node(b0) at (B0) [draw, anchor=west, text width=\sizeD]
       {};
  \node at ($(b0.north)+(0,-0.1)$) [anchor=south] {{\tiny \texttt{prefix}=0}};
  \node(b1) at (B1) [draw, anchor=west, text width=\sizeD]
       {};
  \node at ($(b1.north)+(0,-0.1)$) [anchor=south] {{\tiny \texttt{prefix}=1}};
    
  \node(bucket) at ($(dstate.east)+(\distA,0)+(-0.1,0)$) [anchor=west] {\texttt{Bucket}};

  \coordinate (BS0) at ($(b0.east)+(\distB,0)$);
  \coordinate (BS1) at ($(b1.east)+(\distB,0)$);

  \matrix(bs0) [column sep=-\pgflinewidth, nodes={draw, font=\tiny, inner sep=1, align=center}, anchor=west, text width=0.35cm] at (BS0){
    \node {0001}; & \node {0100};\\
  };

  \draw ($(bs0.north west) + (0.05,-0.05)$) -- ($(bs0.south west) +
  (0.05,0.05)$) -- ($(bs0.south east) + (-0.05,0.05)$) -- ($(bs0.north
  east) + (-0.05,-0.05)$)  -- cycle;

  \matrix(bs1) [column sep=-\pgflinewidth, nodes={draw, font=\tiny, inner sep=1, align=center}, anchor=west, text width=0.35cm] at (BS1){
    \node {1001}; & \node {};\\
  };

  \draw ($(bs1.north west) + (0.05,-0.05)$) -- ($(bs1.south west) +
  (0.05,0.05)$) -- ($(bs1.south east) + (-0.05,0.05)$) -- ($(bs1.north
  east) + (-0.05,-0.05)$)  -- cycle;

  \node(bstate) at ($(bucket.east)+(\distB,0)+(0.2,0)$) [anchor=west] {\texttt{BState}};

  \draw[-latex] (e0.east) -- ($(B0)+(0,0.0)$);
  \draw[-latex] (e1.east) -- ($(B0)+(0,-0.05)$);

  \draw[-latex] (e2.east) -- ($(B1)+(0,0.0)$);
  \draw[-latex] (e3.east) -- ($(B1)+(0,-0.05)$);

  \draw[-latex] (b0.east) -- (BS0);
  \draw[-latex] (b1.east) -- (BS1);

  \coordinate (HT) at ($(e0.north west) - (\distC, 0)$);

  \fill (HT) circle [radius=1pt];
  
  \node (ht) at (HT) [below] {\texttt{ht}};

  \draw[-latex] (HT) -- (e0.north west);

\end{tikzpicture}
    }
    \subfloat[After inserting item $1100$ \label{fig:table_new_item}]{
      \begin{tikzpicture}[
font=\scriptsize,
node distance=0,
text height=1ex, text depth=.25ex,
start chain=1 going below, node distance=-0.15mm]

  \definecolor{mygrey}{RGB}{211,211,211};

  \def\distA{0.7}
  \def\distB{0.5}
  \def\distC{0.25}
  \def\sizeD{0.8cm}

  \node(e0) at (0,0) [draw, on chain=1] {00};
  \node(e1) [draw, on chain=1] {01};
  \node(e2) [draw, on chain=1] {10};
  \node(e3) [draw, on chain=1] {11};

  \node(dstate) at ($(e3.south)+(0,-0.2)$) {\texttt{DState}};

  \coordinate (B0) at ($(e0.east)+(\distA,-0.2)$);
  \coordinate (B1) at ($(e2.east)+(\distA,-0.2)$);

  \node(b0) at (B0) [draw, anchor=west, text width=\sizeD]
       {};
  \node at ($(b0.north)+(0,-0.1)$) [anchor=south] {{\tiny \texttt{prefix}=0}};
  \node(b1) at (B1) [draw, anchor=west, text width=\sizeD]
       {};
  \node at ($(b1.north)+(0,-0.1)$) [anchor=south] {{\tiny \texttt{prefix}=1}};
    
  \node(bucket) at ($(dstate.east)+(\distA,0)+(-0.1,0)$) [anchor=west] {\texttt{Bucket}};

  \coordinate (BS0) at ($(b0.east)+(\distB,0)$);
  \coordinate (BS1) at ($(b1.east)+(\distB,0)$);

  \matrix(bs0) [column sep=-\pgflinewidth, nodes={draw, font=\tiny, inner sep=1, align=center}, anchor=west, text width=0.35cm] at (BS0){
    \node {0001}; & \node {0100};\\
  };

  \draw ($(bs0.north west) + (0.05,-0.05)$) -- ($(bs0.south west) +
  (0.05,0.05)$) -- ($(bs0.south east) + (-0.05,0.05)$) -- ($(bs0.north
  east) + (-0.05,-0.05)$)  -- cycle;

  \matrix(bs1) [column sep=-\pgflinewidth, nodes={draw, font=\tiny, inner sep=1, align=center}, anchor=west, text width=0.35cm] at (BS1){
    \node {1001}; & \node {1100};\\
  };

  \begin{scope}[on background layer]
  \draw[fill=mygrey] ($(bs1.north west) + (0.05,-0.05)$) -- ($(bs1.south west) +
  (0.05,0.05)$) -- ($(bs1.south east) + (-0.05,0.05)$) -- ($(bs1.north
  east) + (-0.05,-0.05)$)  -- cycle;
  \end{scope}

  \node(bstate) at ($(bucket.east)+(\distB,0)+(0.2,0)$) [anchor=west] {\texttt{BState}};

  \draw[-latex] (e0.east) -- ($(B0)+(0,0.0)$);
  \draw[-latex] (e1.east) -- ($(B0)+(0,-0.05)$);

  \draw[-latex] (e2.east) -- ($(B1)+(0,0.0)$);
  \draw[-latex] (e3.east) -- ($(B1)+(0,-0.05)$);

  \draw[-latex] (b0.east) -- (BS0);
  \draw[-latex] (b1.east) -- (BS1);

  \coordinate (HT) at ($(e0.north west) - (\distC, 0)$);

  \fill (HT) circle [radius=1pt];
  
  \node (ht) at (HT) [below] {\texttt{ht}};

  \draw[-latex] (HT) -- (e0.north west);

\end{tikzpicture}
    }
    \newline
    \captionsetup[subfigure]{margin=10pt}
    \subfloat[After bucket splitting (inserting item $0010$)\label{fig:table_new_buckets}]{
        \begin{tikzpicture}[
font=\scriptsize,
node distance=0,
text height=1ex, text depth=.25ex,
start chain=1 going below, node distance=-0.15mm]

  \definecolor{mygrey}{RGB}{211,211,211};

  \def\distA{0.7}
  \def\distB{0.5}
  \def\distC{0.25}
  \def\sizeD{0.8cm}

  \node(e0) at (0,0) [draw, on chain=1, fill=mygrey] {00};
  \node(e1) [draw, on chain=1, fill=mygrey] {01};
  \node(e2) [draw, on chain=1, fill=mygrey] {10};
  \node(e3) [draw, on chain=1, fill=mygrey] {11};

  \begin{scope}[on background layer]
  \draw[white] ($(e0.north)+(0,0.93)$) -- ($(e3.south)+(0,-1)$);
  \end{scope}
  
  \node(dstate) at ($(e3.south)+(0,-0.2)$) {\texttt{DState}};

  \coordinate (B00) at ($(e0.east)+(\distA,0.1)$);
  \coordinate (B01) at ($(e1.east)+(\distA,-0.15)$);

  \coordinate (B1) at ($(e2.east)+(\distA,-0.4)$);

  \node(b00) at (B00) [draw, anchor=west, text width=\sizeD,
    fill=mygrey] {};
  \node at ($(b00.north)+(0,-0.1)$) [anchor=south] {{\tiny \texttt{prefix}=00}};
  \node(b01) at (B01) [draw, anchor=west, text width=\sizeD, fill=mygrey]
       {};
  \node at ($(b01.north)+(0,-0.1)$) [anchor=south] {{\tiny \texttt{prefix}=01}};

  \node(b1) at (B1) [draw, anchor=west, text width=\sizeD]
       {};
  \node at ($(b1.north)+(0,-0.1)$) [anchor=south] {{\tiny
      \texttt{prefix}=1}};
  
  \node(bucket) at ($(dstate.east)+(\distA,0)+(-0.1,0)$) [anchor=west] {\texttt{Bucket}};

  \coordinate (BS00) at ($(b00.east)+(\distB,0)$);
  \coordinate (BS01) at ($(b01.east)+(\distB,0)$);

  \coordinate (BS1) at ($(b1.east)+(\distB,0)$);

  \matrix(bs00) [column sep=-\pgflinewidth, nodes={draw, font=\tiny, inner sep=1, align=center}, anchor=west, text width=0.35cm] at (BS00){
    \node {0001}; & \node {0010};\\
  };

  \begin{scope}[on background layer]
  \draw[fill=mygrey] ($(bs00.north west) + (0.05,-0.05)$) --
  ($(bs00.south west) + (0.05,0.05)$) -- ($(bs00.south east) +
  (-0.05,0.05)$) -- ($(bs00.north east) + (-0.05,-0.05)$) -- cycle;
  \end{scope}

  \matrix(bs01) [column sep=-\pgflinewidth, nodes={draw, font=\tiny,
      inner sep=1, align=center}, anchor=west, text width=0.35cm] at (BS01){
    \node {0100}; & \node {};\\
  };
  
  \begin{scope}[on background layer]
  \draw[fill=mygrey] ($(bs01.north west) + (0.05,-0.05)$) -- ($(bs01.south west) +
  (0.05,0.05)$) -- ($(bs01.south east) + (-0.05,0.05)$) -- ($(bs01.north
  east) + (-0.05,-0.05)$)  -- cycle;
  \end{scope}

  \matrix(bs1) [column sep=-\pgflinewidth, nodes={draw, font=\tiny,inner sep=1, align=center}, anchor=west, text width=0.35cm] at (BS1){
    \node {1001}; & \node {1100};\\
  };

  \draw ($(bs1.north west) + (0.05,-0.05)$) -- ($(bs1.south west) +
  (0.05,0.05)$) -- ($(bs1.south east) + (-0.05,0.05)$) -- ($(bs1.north
  east) + (-0.05,-0.05)$)  -- cycle;

  \node(bstate) at ($(bucket.east)+(\distB,0)+(0.2,0)$) [anchor=west] {\texttt{BState}};

  \draw[-latex] (e0.east) -- ($(B00)+(0,0.0)$);
  \draw[-latex] (e1.east) -- ($(B01)+(0,-0.05)$);

  \draw[-latex] (e2.east) -- ($(B1)+(0,0.0)$);
  \draw[-latex] (e3.east) -- ($(B1)+(0,-0.05)$);

  \draw[-latex] (b00.east) -- (BS00);
  \draw[-latex] (b01.east) -- (BS01);
  \draw[-latex] (b1.east) -- (BS1);

  \coordinate (HT) at ($(e0.north west) - (\distC, 0)$);

  \fill (HT) circle [radius=1pt];
  
  \node (ht) at (HT) [below] {\texttt{ht}};

  \draw[-latex] (HT) -- (e0.north west);

\end{tikzpicture}
    }
    \subfloat[After directory resizing (inserting item $0000$)\label{fig:table_new_directory}]{
        \begin{tikzpicture}[
font=\scriptsize,
node distance=0,
text height=1ex, text depth=.25ex,
start chain=1 going below, node distance=-0.15mm]

  \definecolor{mygrey}{RGB}{211,211,211};

  \def\distA{0.7}
  \def\distB{0.5}
  \def\distC{0.25}
  \def\sizeD{0.8cm}

  \node(e0) at (0,0) [draw, on chain=1, fill=mygrey] {000};
  \node(e1) [draw, on chain=1, fill=mygrey] {001};
  \node(e2) [draw, on chain=1, fill=mygrey] {010};
  \node(e3) [draw, on chain=1, fill=mygrey] {011};
  \node(e4) at (0,0) [draw, on chain=1, fill=mygrey] {100};
  \node(e5) [draw, on chain=1, fill=mygrey] {101};
  \node(e6) [draw, on chain=1, fill=mygrey] {110};
  \node(e7) [draw, on chain=1, fill=mygrey] {111};

  \node(dstate) at ($(e7.south)+(0,-0.2)$) {\texttt{DState}};

  \coordinate (B000) at ($(e0.east)+(\distA,-0.15)$);
  \coordinate (B001) at ($(e1.east)+(\distA,-0.45)$);
  \coordinate (B01) at ($(e2.east)+(\distA,-0.85)$);
  \coordinate (B1) at ($(e5.east)+(\distA,-0.5)$);

  \node(b000) at (B000) [draw, anchor=west, text width=\sizeD, fill=mygrey]
       {};
  \node at ($(b000.north)+(0,-0.1)$) [anchor=south] {{\tiny \texttt{prefix}=000}};
  \node(b001) at (B001) [draw, anchor=west, text width=\sizeD, fill=mygrey]
       {};
  \node at ($(b001.north)+(0,-0.1)$) [anchor=south] {{\tiny \texttt{prefix}=001}};
  \node(b01) at (B01) [draw, anchor=west, text width=\sizeD]
       {};
  \node at ($(b01.north)+(0,-0.1)$) [anchor=south] {{\tiny \texttt{prefix}=01}};

  \node(b1) at (B1) [draw, anchor=west, text width=\sizeD]
       {};
  \node at ($(b1.north)+(0,-0.1)$) [anchor=south] {{\tiny \texttt{prefix}=1}};
  
  \node(bucket) at ($(dstate.east)+(\distA,0)+(-0.1,0)$) [anchor=west] {\texttt{Bucket}};

  \coordinate (BS000) at ($(b000.east)+(\distB,0)$);
  \coordinate (BS001) at ($(b001.east)+(\distB,0)$);
  \coordinate (BS01) at ($(b01.east)+(\distB,0)$);
  \coordinate (BS1) at ($(b1.east)+(\distB,0)$);

  \matrix(bs000) [column sep=-\pgflinewidth, nodes={draw, font=\tiny, inner sep=1, align=center}, anchor=west, text width=0.35cm] at (BS000){
    \node {0001}; & \node {0000};\\
  };

  \begin{scope}[on background layer]
  \draw[fill=mygrey] ($(bs000.north west) + (0.05,-0.05)$) --
  ($(bs000.south west) + (0.05,0.05)$) -- ($(bs000.south east) +
  (-0.05,0.05)$) -- ($(bs000.north east) + (-0.05,-0.05)$) -- cycle;
  \end{scope}

  \matrix(bs001) [column sep=-\pgflinewidth, nodes={draw, font=\tiny,
      inner sep=1, align=center}, anchor=west, text width=0.35cm] at (BS001){
    \node {0010}; & \node {};\\
  };
  
  \begin{scope}[on background layer]
  \draw[fill=mygrey] ($(bs001.north west) + (0.05,-0.05)$) -- ($(bs001.south west) +
  (0.05,0.05)$) -- ($(bs001.south east) + (-0.05,0.05)$) -- ($(bs001.north
  east) + (-0.05,-0.05)$)  -- cycle;
  \end{scope}

  \matrix(bs01) [column sep=-\pgflinewidth, nodes={draw, font=\tiny,inner sep=1, align=center}, anchor=west, text width=0.35cm] at (BS01){
    \node {0100}; & \node {};\\
  };

  \draw ($(bs01.north west) + (0.05,-0.05)$) -- ($(bs01.south west) +
  (0.05,0.05)$) -- ($(bs01.south east) + (-0.05,0.05)$) -- ($(bs01.north
  east) + (-0.05,-0.05)$)  -- cycle;

  \matrix(bs1) [column sep=-\pgflinewidth, nodes={draw, font=\tiny,inner sep=1, align=center}, anchor=west, text width=0.35cm] at (BS1){
    \node {1001}; & \node {1100};\\
  };

  \draw ($(bs1.north west) + (0.05,-0.05)$) -- ($(bs1.south west) +
  (0.05,0.05)$) -- ($(bs1.south east) + (-0.05,0.05)$) -- ($(bs1.north
  east) + (-0.05,-0.05)$)  -- cycle;

  \node(bstate) at ($(bucket.east)+(\distB,0)+(0.2,0)$) [anchor=west] {\texttt{BState}};

  \draw[-latex] (e0.east) -- ($(B000)+(0,0.0)$);
  \draw[-latex] (e1.east) -- ($(B001)+(0,-0.05)$);

  \draw[-latex] (e2.east) -- ($(B01)+(0,0.0)$);
  \draw[-latex] (e3.east) -- ($(B01)+(0,-0.05)$);

  \draw[-latex] (b000.east) -- (BS000);
  \draw[-latex] (b001.east) -- (BS001);
  \draw[-latex] (b01.east) -- (BS01);

  \draw[-latex] (e4.east) -- ($(B1)+(0,0.1)$);
  \draw[-latex] (e5.east) -- ($(B1)+(0,0.025)$);
  \draw[-latex] (e6.east) -- ($(B1)+(0,-0.025)$);
  \draw[-latex] (e7.east) -- ($(B1)+(0,-0.1)$);

  \draw[-latex] (b1.east) -- (BS1);

  \coordinate (HT) at ($(e0.north west) - (\distC, 0)$);

  \fill (HT) circle [radius=1pt];
  
  \node (ht) at (HT) [below] {\texttt{ht}};

  \draw[-latex] (HT) -- (e0.north west);

\end{tikzpicture}
    }
    \caption{Operations on the wait-free extendible hash table. (The
      new objects created during an operation are grayed) \label{fig:table_work}}
  \end{center}
\end{figure}
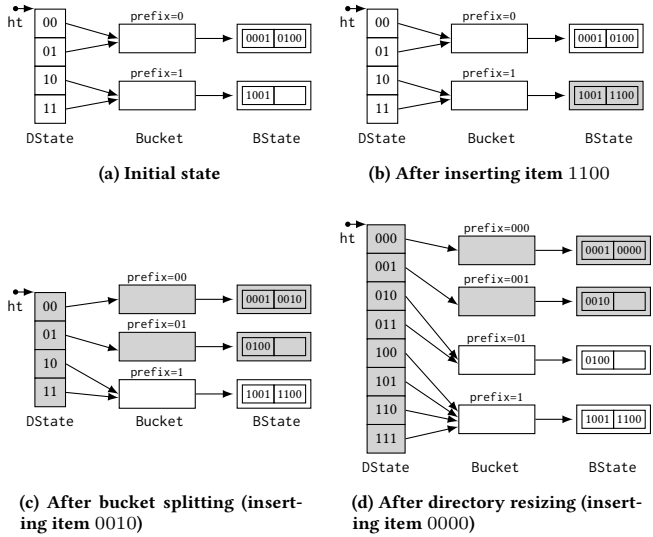

Figure~\ref{fig:table_work} presents the structure of our hash table
and describes how it evolves when update operations are executed.
The main challenge when designing a resizable hash table 
is allowing operations of the hash table to be
executed in parallel with resizing actions. To this end, our algorithm
uses two levels of indirection between the \texttt{DState} object that
implements the directory and the \texttt{BState} objects that store
the items of buckets (see Figure~\ref{fig:table_init}). 
In a \texttt{BState} object, items are stored
in a fixed size array (in Figure~\ref{fig:table_work}, this size is $2$).
The data records used to implement our hash table are presented in
Figure~\ref{fig:bpsim_struct}.  To distinguish a variable that stores a
reference to an object, we add the suffix \texttt{\_p} to its type.

To insert a new element into a non-full bucket in a wait-free manner, a
thread creates a local copy of the corresponding \texttt{BState}. It
applies its operation on this local copy and uses a
\texttt{CompareAndSwap} (\CAS) operation to attempt to update the \texttt{BState}
pointer in the corresponding bucket to point to its local
\texttt{BState} object (see Figure~\ref{fig:table_new_item}).
If the corresponding bucket is full, 
it will be replaced by two new buckets to complete the \INSERT. 
This case is illustrated in Figure~\ref{fig:table_new_buckets}. 
The directory, i.e., the \texttt{DState} object, must be updated to
store references to the new buckets. To perform the update on the directory state
in a wait-free manner, a thread first creates a local copy of the
currently active \texttt{DState} object (pointed to by the shared variable
\texttt{ht}). Note that this copy contains pointers
to the existing buckets. The thread can then create
the new buckets from the full bucket (buckets $00$ and $01$ from bucket $0$ in
Figure~\ref{fig:table_new_buckets}), update its local \texttt{DState} object
accordingly, and finally, try to make its local \texttt{DState} object the
active \texttt{DState} object by updating \texttt{ht} using \CAS.

The algorithm has to ensure that update operations are not lost if
they run concurrently with a resizing action that requires replacing
the current \texttt{DState} object. 
Thus, our algorithm has to handle the following two cases that might
arise during the creation of a new \texttt{DState} object:
(i) no update should be lost if a pointer to a bucket is replaced with pointers to new buckets
in the new \texttt{DState} object,
and (ii) no update targeting the non-full buckets should be lost, 
i.e., the non-full buckets must still be referenced by the new \texttt{DState} object. 
To prevent the appearance of the first case, 
our algorithm ensures that (1) a bucket is split only if it is full and
(2) no update operation (not even \DELETE) is executed on a full bucket. 
To prevent the  second case from occurring, 
our algorithm employs two levels of indirection 
between a \texttt{DState} object and the
\texttt{BState} objects. Assume that a \texttt{BState} object is updated
as shown in Figure~\ref{fig:table_new_item}, 
while the \texttt{DState} object pointing to this bucket is also 
updated (as shown Figure~\ref{fig:table_new_buckets}).
In this case, the new \texttt{DState} object will still point 
to the previously existing \texttt{Bucket} objects that are not split
(i.e., that are not full). Thus, even if
the \texttt{BState} reference stored in a \texttt{Bucket} changes, 
it will still be accessible through the new \texttt{DState} object.

Resizing the directory is required when its depth becomes smaller 
than the depth of new buckets that are created during an \INSERT\ 
operating on a full bucket. In this case, after creating a
local copy of the currently active \texttt{DState} object
(i.e., after copying all pointers to buckets locally), 
the thread executing the \INSERT\ must create a new \texttt{DState} object
of larger depth, and copy the existing bucket references into it.
Figure~\ref{fig:table_new_directory} illustrates this case.

In the following sections, we provide the details of our algorithm. 
We discuss how update operations on \texttt{DState} and \texttt{BState} objects 
are implemented in a wait-free manner, 
as well as how they are synchronized to ensure linearizability.

\subsection{Data structures for the resizable hash table implementation \label{sub:ds}}

Recall that the data records used to implement our hash table are presented in
Figure~\ref{fig:bpsim_struct}. Part of the \texttt{Bucket} state is stored in 
a \texttt{BState} record. This part is copied by
every thread that wants to apply an atomic update on the bucket.
Each \texttt{Bucket} record, as well as each \texttt{BState} record includes a bit vector of
size $n$ (called \texttt{toggle} and \texttt{applied}, respectively) that are used to
efficiently track the pending operations on a bucket. A
\texttt{BState} record also stores an array, called \texttt{Result},
that is used to store the results of active operations.
(We describe how these fields are used in more detail in the subsequent sections.)
The fixed size
array\footnote{The size of the array equals the maximum number of
  elements a bucket can store. To look for a key, a thread has to
  iterate over all elements currently stored in the array.} used to
store the items associated with a \texttt{BState} is called
$items$. In the following, we use the notation \emph{items[key]} to
set or get the value associated with one key.
The hash table is represented by a record of type \texttt{DState} that
is composed of an array of references to elements of type
\texttt{Bucket} and a variable \texttt{depth} storing the depth of
  the directory.  

The shared variables and the
private persistent variables for each thread are provided in
Figure~\ref{fig:variables}.
Each operation initiated by a process $i$ is announced in
element $i$ of the shared array \emph{help}. 
A per-thread sequence number
($opSeqnum_i$) is introduced but it will only become useful when
discussing resizing in Section~\ref{sub:extendible}.

\begin{figure}[t]
\begin{lstlisting}[style=2col]
struct Operation:
  $type$   : {$INS$, $DEL$}
  $key$    : integer
  $value$ : integer
  $seqnum$ : integer

struct Result:
  $status$ : {$TRUE$, $FALSE$, $FAIL$}
  $seqnum$ : integer

struct DState:
  $depth$  : integer
  $dir[2^{depth}]$ : Bucket_p
  
struct Bucket:
  $prefix$  : bit-string
  $depth$   : integer
  $state$   : BState_p
  $toggle[n]$  : boolean


  
struct BState:
  $items$   : set of key-value pairs
  $applied[n]$ : boolean
  $results[n]$ : Result
\end{lstlisting}
\caption{Data structures definitions (for $n$ threads).\label{fig:bpsim_struct}}
\end{figure}

\begin{figure}[t]
\begin{lstlisting}[style=2col]
Shared variables:
  $ht$ : DState_p
  $help[n]$  : Operation_p
Persistent private variable:
    $opSeqnum_i$  : integer
\end{lstlisting}
\caption{Variables for thread $T_i$.\label{fig:variables}}
\end{figure}

\subsection{The case of no resizing}
\label{sub:basic}

This section describes how \INSERT\ and
\LOOKUP\ work when no resizing is required.
Figure~\ref{fig:ht_interface} provides code for \INSERT\ and
\LOOKUP. \DELETE\ is implemented in the same way as \INSERT. 

\begin{figure}[t]
\begin{lstlisting}[style=1col]
def Lookup(integer key): (boolean, integer) ~\label{line:lookup1}~
    htl $\leftarrow$ ht $\star$ ~\label{line:htref1}~ 
    bs $\leftarrow$ htl.dir[Prefix(key, htl.depth)].bstate ~\label{line:lookupbucket}~
    return ((key $\in$ bs.items) ? (TRUE, bs.items[key]) : (FALSE, -1))~\label{line:lookup2}~

def Insert(integer key, integer value): boolean
    $opSeqnum_i$ $\leftarrow$ $opSeqnum_i$+1 $\star$~\label{line:seqnum}~
    help[i] $\leftarrow$ new Operation($INS$, key, value, $opSeqnum_i$) ~\label{line:reqhelp}~
    htl $\leftarrow$ ht $\star$ ~\label{line:htref2}~
    ApplyWFOp( htl.dir[ Prefix(key, htl.depth) ] ) ~\label{line:callApplyExt}~
    htl $\leftarrow$ ht $\star$ ~\label{line:htref3}~
    if htl.dir[ Prefix(key, htl.depth) ].state.results[i].seqnum != $opSeqnum_i$: $\star$ ~\label{line:checkinsertfail}~
        ResizeWF() $\star$ ~\label{line:callresize}~
    htl $\leftarrow$ ht $\star$ ~\label{line:htref4}~
    return htl.dir[ Prefix(key, htl.depth) ].state.results[i].status ~\label{line:getresult}~
    
def ApplyWFOp(Bucket_p b):
    Flip(b.toggle, i) ~\label{line:toggleset}~# change the value of bit i 
    for k in 1..2: ~\label{line:loopstart}~
        oldb $\leftarrow$ b.state ~\label{line:copyref}~
        newb $\leftarrow$ New BState(oldb) # create a copy of object oldb ~\label{line:bstate}~
        t $\leftarrow$ b.toggle ~\label{line:copytoggle}~
        for j in [ tid for tid in 1..n if t[tid] != newb.applied[tid] ]: ~\label{line:testset}~
            if newb.results[j].seqnum < help[j].seqnum: $\star$ ~\label{line:applycheck}~
                newb.results[j].status $\leftarrow$ ExecOnBucket(newb, help[j]) ~\label{line:result1}~
                if newb.results[j].status is not FAIL: $\star$
                    newb.results[j].seqnum $\leftarrow$ help[j].seqnum $\star$ ~\label{line:result2}~
        newb.applied $\leftarrow$ t ~\label{line:appliedset}~
        CAS( b.state, oldb, newb) ~\label{line:CAS}~

def ExecOnBucket(BState_p b, Operation_p op): status ~\label{line:startexec}~ 
    if b.items is full: ~\label{line:teststatefull}~
        return $FAIL$ ~\label{line:returnfail}~
    else:
        exist $\leftarrow$ (op.key $\in$ b.items)
        if op.type == $INS$:
            b.items[op.key] $\leftarrow$ op.value # updates or inserts
            return !exist
        else:
            delete b.items[op.key] # does nothing if key not present
            return exist ~\label{line:endexec}~
\end{lstlisting}
\caption{Code of \INSERT\ and \LOOKUP\ operations
(lines marked with the $\star$ symbol are required only to implement resizing).\label{fig:ht_interface}}
\end{figure}

As described in Section~\ref{sub:nutshell}, to insert an item 
in a non-full bucket $b$, a thread must update the corresponding
\texttt{BState}. This is done in \texttt{ApplyWFOp()}. 
Note that the thread executes other pending updates 
on the bucket (in addition to its own), leveraging the core idea of \PSIM. 
Specifically, the thread first announces its operations
(lines~\ref{line:reqhelp} and \ref{line:toggleset}). Then it tries to
apply its operation, as well as all other pending operations for
bucket $b$, on a local copy $newb$ of the bucket state (lines
\ref{line:copyref}-\ref{line:appliedset}). Finally, it executes a
\CAS\ to make this updated state the new value of $b.state$
(line~\ref{line:CAS}). When a thread executes an operation on behalf
of another thread, it stores the result of this operation in
$newb.results$ so that the other thread can find there the result of
its operation (line~\ref{line:getresult}).

To identify pending updates on a bucket, the \texttt{toggle} vector of
the \texttt{Bucket} and the \texttt{applied} vector of the
corresponding \texttt{BState} are used. To announce an update
operation $op$ on a bucket $b$, a thread $T_i$ flips its bit in
$b.toggle$ (line~\ref{line:toggleset}) in an atomic way.
Hence, if a bit $j$ has different values 
in $b.toggle$ and $b.state.applied$, then thread $T_j$ has
a pending operation.  Note that $T_i$
announces $op$ in $help[i]$ before flipping its bit in $b.toggle$.

After $T_i$ has executed the body of the loop on lines
\ref{line:loopstart}-\ref{line:CAS} twice, it is guaranteed that the
operation of thread $T_i$ has been executed~\cite{Fatourou2013}: If
both $T_i$'s \CAS\ operations fail (line~\ref{line:CAS}), it means that another thread $T_j$
has run an iteration of this loop and has executed a successful \CAS\ 
between $T_i$'s first and second \CAS\ step. Since an iteration executes all pending
operations and thread $T_i$ had registered its own operation before
executing its first \CAS\ step (line~\ref{line:toggleset}), 
$T_j$  has applied the operation of thread $T_i$. 
So, after executing it second \CAS\ step, 
$T_i$ can read the result of its operation 
in the corresponding entry of the $result$ array of the active
\texttt{BState} (line~\ref{line:getresult}). 
The function \texttt{ExecOnBucket()} (lines
\ref{line:startexec}-\ref{line:endexec}) is called to (sequentially)
execute an update operation on a local copy of a bucket state.

Complying with design rule (A), \LOOKUP\ operations
(line \ref{line:lookup1}) do not require any synchronization, i.e.,
they execute a code that is exactly the same as their sequential code 
independently of whether the buckets 
they are directed at are being resized when they are executed. 
Doing so does not violate linearizability because when a
thread makes a copy of the current \texttt{BState} to
apply an update to it, it copies the set of items as part of this
state. Therefore, the bucket states are immutable,
allowing to access them safely without synchronization. 
Performing this copy has a cost for update operations. 
However, it allows for very efficient \LOOKUP\ operations 
which are the most frequent operations performed on a hash table.

To access a bucket (e.g., line \ref{line:lookupbucket}), 
a thread first stores a copy of $ht$ in local variable $htl$
(lines~\ref{line:htref1},~\ref{line:htref2},~\ref{line:htref3}, and~\ref{line:htref4}). 
This is needed because $ht.dir$ and $ht.depth$ should refer to
the same \texttt{DState}. This might not be the case without 
copying the reference stored in $ht$ locally, given that $ht$ may be concurrently updated by resizing actions.

\subsection{Resizing the hash table}
\label{sub:extendible}

Figure~\ref{fig:ht_interface_ext} describes the operations
  required to resize the hash table, i.e., to split 
each of the full buckets into two buckets, as well as to increase the
  size of the directory. We defer the discussion of bucket merging to
  Section \ref{sub:merging}.

\begin{figure}[t]
\begin{lstlisting}[style=1col]
def SplitBucket(Bucket_p b): (Bucket_p, Bucket_p)
    b0 $\leftarrow$ new Bucket(b) # copy of bucket b~\label{line:newBuck0}~
    b0.depth $\leftarrow$ b.depth + 1 
    b0.prefix $\leftarrow$ b.prefix << 1 # bucket with prefix $B_1B_2B_3...0$
    b0.state $\leftarrow$ new BState() # new empty BState
    b0.state.results $\leftarrow$ b.state.results #copy if the results array of b ~\label{line:copyresult}~
    b0.state.applied $\leftarrow$ b0.toggle ~\label{line:initapplied}~
    b1 $\leftarrow$ new Bucket(b0) ~\label{line:newBuck1}~
    b1.prefix $\leftarrow$ b0.prefix + 1 # bucket with prefix $B_1B_2B_3...1$  
    for (k,v) in b.state.items: # insert the key-value pairs in new BStates ~\label{line:startsplititems}~
        if Prefix(k, b0.depth) == b0.prefix:
            b0.state.items[k] $\leftarrow$ v
        else:
            b1.state.items[k] $\leftarrow$ v ~\label{line:endsplititems}~
    return b0, b1

def DirectoryUpdate(DState_p d, Bucket_p blist[]):
    for b in blist:
        if b.depth > d.depth: # doubling directory size is required ~\label{line:testdepth}~
            alloc d.dir with size $2^{d.depth + 1}$ ~\label{line:realloc}~
            copy all Bucket_p from previous dir to new d.dir ~\label{line:copybuckref}~
            d.depth $\leftarrow$ d.depth + 1
        #compute the set of directory entries that should point to b
        entries $\leftarrow$ {e for e in 1..$2^{d.depth}$ if Prefix(e, b.depth) = b.prefix} ~\label{line:startentryup}~
        for e in entries: # insert b in all required entries of d.dir
            d.dir[e] $\leftarrow$ b ~\label{line:endentryup}~

def ApplyPendingResize(DState_p d, Bucket_p bFull):
    for j in 1..n: ~\label{line:iterhelpapp}~
        if Prefix(help[j].op.key, bFull.depth) == bFull.prefix: ~\label{line:testprefixapp}~
            if bFull.state.results[j].seqnum < help[j].seqnum: ~\label{line:testappliedapp}~
                # help[j] is a pending request that was applying to bFull
                # bDest is the destination of help[j] in the current state
                bDest = d.dir[ Prefix(help[j].op.key, d.depth) ] ~\label{line:selectbUp}~
                while bDest is full: ~\label{line:startsplitloop}~
                    (b0,b1) $\leftarrow$ SplitBucket(bDest) ~\label{line:callsplit}~
                    DirectoryUpdate(d, (b0, b1)) ~\label{line:calldirup}~
                    #update bDest based on the new state of the directory
                    bDest $\leftarrow$ d.dir[ Prefix(help[j].op.key, d.depth) ] ~\label{line:selectnewbUp}~
                bDest.state.results[j].status $\leftarrow$ ExecOnBucket(bDest, help[j]) ~\label{line:execending}~
                bDest.state.results[j].seqnum $\leftarrow$ help[j].seqnum ~\label{line:saveseqnumpending}~
    
def ResizeWF(): ~\label{line:resizeWF}~
    for k in 1..2:
        oldD $\leftarrow$ ht
        newD $\leftarrow$ new DState(oldD) ~\label{line:dstatecopy}~# copy of object oldD
        for j in 1..n: ~\label{line:iterhelp}~ # iterates over operations in help
            b $\leftarrow$ newD.dir[ Prefix(help[j].op.key, newD.depth) ]
            if b is full and b.state.results[j].seqnum < help[j].seqnum: ~\label{line:pendingcheck}~
                ApplyPendingResize(newD, b) ~\label{line:callpending}~
        CAS(ht, oldD, newD) ~\label{line:CAS2}~
\end{lstlisting}
\caption{Code for the Resizable Hash Table.\label{fig:ht_interface_ext}}
\end{figure}

\noindent{\bf Updating the directory.} When an \INSERT\ operation
fails because a bucket is full, the calling thread runs the
\texttt{ResizeWF()} function (line~\ref{line:callresize}). 
The algorithm executed by \texttt{ResizeWF()} follows the same basic principle  
as that executed by \texttt{ApplyWFOp()}.  
All threads willing to apply a resize action (here
splitting a full bucket) will run two iterations of a loop where they
will first make a local copy of the directory
(line~\ref{line:dstatecopy}), then apply the required modifications on
their local copy, and finally try to make their local copy the active
state of the directory using a \CAS\ instruction
(line~\ref{line:CAS2}). However, there is no need to use bit vectors to
determine the actions to be executed on the directory: 
all such actions will be resizing actions.

Making \texttt{ApplyWFOp()} and \texttt{ResizeWF()} wait free is not
enough to ensure that \INSERT\ is wait free. We need to also
ensure that a thread calls these functions a bounded number
of times to complete an \INSERT. 
To this end, special care should
be given to \emph{pending resize actions} and pending update
operations during resizing. A {\em pending resize action} is a resize
action that must be performed in order for a pending update on a
bucket to complete. 
Specifically, 
 (i) a thread running a resize
action has to execute all pending resize actions on the current state
of the hash table, and (ii) when creating a new bucket, a thread has to
execute all pending update operations on the newly created bucket. 
We explain the necessity of each of these two special actions in the next two paragraphs.

Consider the case where two threads, $T_1$ and $T_2$, need to execute
\texttt{ResizeWF()} to complete an \INSERT\ operation each, 
on distinct buckets.
Assume that $T_1$ needs to split
bucket $b_a$ and $T_2$ needs to split bucket $b_c$ to complete their operations. 
If $T_1$ tries to update the directory in parallel with $T_2$,
without executing the split of bucket $b_c$, there is a chance that
only $T_1$ will manage to make its new computed state active using
\texttt{CAS} (line \ref{line:CAS2}), requiring $T_2$ to run again its
resize action. The second time, a thread $T_3$ could be trying to
split a bucket $b_d$ in parallel with $T_2$ splitting $b_c$, and make
$T_2$ fail again. As such a scenario could keep happening, 
a thread that executes \texttt{ResizeWF()} executes all pending resize actions 
in the current state of the directory in order to ensure wait-freedom.

Consider now the following scenario where multiple threads run
\INSERT\ on a single bucket that is to be split. Assume that a thread $T_1$
applies an \INSERT\ on a full bucket $b$ and so it calls
\texttt{ResizeWF()}. At this time, it experiences some delay and in
the meantime another thread $T_2$ executes resizing actions
and replaces the full bucket $b$ with two new buckets. Thread $T_1$
would now have to apply its \INSERT\ operation on one of the newly
created buckets.  However, suppose that other threads applied
\INSERT\ operations in the meantime, so that this new bucket is full
again.  In this case, the operation of thread $T_1$ would fail again
and $T_1$ would have to execute another resize action using
\texttt{ResizeWF()}, with the risk that the same thing would happen
again. To prevent such a scenario, 
a thread creating a new bucket executes
all pending updates directed to this bucket.

\noindent
{\bf Executing each operation exactly once.} Since \INSERT\ operations
may be executed through two different paths, namely during 
the execution of \texttt{ApplyWFOp()} (line \ref{line:result1}) 
or during the execution of 
\texttt{ApplyPendingResize()} (line \ref{line:execending}) which is called by
\texttt{ResizeWF()}  (line \ref{line:callpending}), we need a way to
figure out when an operation has already been executed or is still
pending. To this end, we use a per-thread sequence number
$opSeqnum$. A thread $T_i$ tags each update operation it executes with a
distinct sequence number (line \ref{line:reqhelp}). When an operation
of $T_i$ has been successfully executed on a bucket $b$, its
sequence number is copied into the entry $i$ of the \texttt{results}
array associated with $b.state$, i.e., \texttt{results[i].seqnum} is the
sequence number of the last update executed by $T_i$ on $b$.

The sequence number of the operations announced in \texttt{help[j]}
(for some $j$) targeting a bucket $b$ is compared to the
corresponding entry \texttt{b.state.results[j].seqnum} to determine
whether the operation is pending. This test is made both in
\texttt{ApplyWFOp()} (line \ref{line:applycheck}) and in
\texttt{ApplyPendingResize()} (line \ref{line:testappliedapp}) to
decide which operations to execute on the bucket.

Note that in \texttt{ApplyWFOp()}, a thread cannot solely rely on the
toggle vector of the bucket to decide which operations to execute
(line \ref{line:testset}). Indeed, since a thread starts an update
operation by registering its operation in the \texttt{help} array
(line \ref{line:reqhelp}), there is a chance that its operation would
be executed during a resizing action even before it calls
\texttt{ApplyWFOp()} and flips its bit in the toggle vector (line
\ref{line:toggleset}). As such, the only safe way to identify pending
updates is using the sequence numbers. The toggle vector associated
with each bucket is used to improve performance: It allows to identify
very fast potentially pending updates without having to read the whole
\texttt{help} array in \texttt{ApplyWFOp()}.

\noindent
{\bf Detailed description of the resizing algorithm.}  We now describe
the \texttt{ResizeWF()} function in detail. After making a local copy
of the directory state, a thread iterates over the \texttt{help} array
to find \emph{pending resize actions} (lines
\ref{line:iterhelp}-\ref{line:pendingcheck}); specifically, the thread
looks in {\tt help} for pending update operations that apply to a full
bucket.  For each pending resize action, the thread calls
\texttt{ApplyPendingResize()} with the corresponding bucket, $bFull$, 
as parameter.

In \texttt{ApplyPendingResize()}, a thread should run all pending
updates that are to be applied to bucket $bFull$. To this end, it
iterates over the help array to find the pending operations that apply
to $bFull$ (lines
\ref{line:iterhelpapp}-\ref{line:testappliedapp}). For such an
operation, it selects the bucket $bDest$ on which the operation should
be applied in the current local directory state (line
\ref{line:selectbUp}). As long as $bDest$ is full, it splits $bDest$
into two buckets $b0$ and $b1$ (call to \texttt{SplitBucket()} line
\ref{line:callsplit}), updates the directory with the two new buckets
calling \texttt{DirectoryUpdate()} (line \ref{line:calldirup}), and
updates $bDest$ based on the new state of the directory (line
\ref{line:selectnewbUp}). Finally, it executes the pending operation
on bucket $bDest$ and updates the sequence number in
$bDest.state.results$ accordingly (lines
\ref{line:execending}-\ref{line:saveseqnumpending}).

We remark that $bDest$ can be different from $bFull$ when executing
line \ref{line:selectbUp}.  The first operation applied on $bFull$
will split $bFull$ and replace it by non-full buckets. Hence, for
other operations targeting $bFull$, $bDest$ should be set to point
directly to the appropriate newly created bucket. Note also that
buckets might have to be split several times before completing a
single \INSERT\ operation (lines
\ref{line:startsplitloop}-\ref{line:selectnewbUp}). The reason is that
when one full bucket $b$ is split into two new buckets $b0$ and $b1$,
there is a chance that all items stored in $b$ should be moved to just
one of the two new buckets. For instance, in
Figure~\ref{fig:hash_after}, when splitting bucket $010$, all items
will be stored in bucket $0100$, which thus should be split again.

Function \texttt{SplitBucket()} takes, as a parameter, a reference to
a bucket $b$ of depth $D$ with prefix "$B_1B_2\ldots B_D$" and returns
references to two new buckets $b_0$ and $b_1$ of depth $D+1$, with
prefixes "$B_1B_2\ldots B_D0$" and "$B_1B_2\ldots B_D1$".  Key-value
pairs stored in the \texttt{BState} of $b$ are copied to the
\texttt{BState} of $b0$ and $b1$ based on their prefix (lines
\ref{line:startsplititems}-\ref{line:endsplititems}). We recall that
since $b$ is full, the algorithm guaranties that its \texttt{BState}
is immutable.  Note also that $b.state.results$ is copied in $b0$'s
and $b1$'s state (line \ref{line:copyresult}) so that the results of
operations previously applied to $b$ are also available through $b0$
and $b1$. Finally, for each new bucket, its \texttt{applied} vector is
initialized to be equal to its \texttt{toggle} vector (line
\ref{line:initapplied}): since all pending updates on the
bucket are executed by the \texttt{ApplyPendingResize()} function, and
since the bucket is not visible to other threads as long as the
\CAS\ instruction that updates the active directory has not been
executed (line \ref{line:CAS2}), its toggle vector should reflect no
pending operations.

\texttt{DirectoryUpdate()} inserts in the directory the new buckets
generated by \texttt{SplitBucket()}. The function doubles the size of
the directory if needed, and stores the references of the new buckets
in the appropriate entries of the directory.  Although we do not
detail all steps to be executed to double the size of the directory,
the two main steps are: (i) allocating a new array corresponding to
the new size of the directory (line \ref{line:realloc}), and (ii)
copying all the bucket references from the previous array to the
appropriate entries of the new array (line
\ref{line:copybuckref}). Note that a new bucket might have to be
inserted in more than one entry if the depth of the directory if
bigger than the depth of the bucket (lines
\ref{line:startentryup}-\ref{line:endentryup}).

\noindent
{\bf Avoiding losing updates.}  Our algorithm allows update operations
on non-full buckets to run concurrently with resize actions. As
pointed in Section~\ref{sub:nutshell}, the two levels of indirection
between the \texttt{DState} objects and the \texttt{BState} objects
storing items ensure that updates are not lost in this case. Indeed,
when a thread creates a new directory state during resizing, it copies
the references of all existing buckets (line \ref{line:dstatecopy}) in
the new state. Since applying an update on a bucket $b$ only modifies
the \texttt{BState} reference stored in $b.state$ (line
\ref{line:CAS}), the newly created directory state will allow
accessing the concurrently updated \texttt{BState} objects.

Note that since resizing requires full buckets to be immutable to
avoid losing updates when running \texttt{SplitBucket()},
\DELETE\ operations must not be executed on a full bucket (lines
\ref{line:teststatefull}-\ref{line:returnfail}).

\noindent
    {\bf Compliance with the design rules.}  
The described algorithm complies with the two design rules we have
previously defined. (A) \LOOKUP\ operations are executed without ever
requiring any synchronization. The \LOOKUP\ implementation is
equivalent to a sequential implementation. (B) When no resize action
is running, an update operation is executed by the instance of
\PSIM\ of the corresponding bucket (function \texttt{ApplyWFOp()})
fully in parallel with operations applying to other buckets. The use
of per-thread sequence numbers does not impair execution parallelism.

\subsection{Merging buckets and shrinking the directory \label{sub:merging}}

We now provide a high level description of how our implementation
copes with merging and shrinking.
Merging buckets and shrinking the directory should both be run through
the \texttt{ResizeWF()} function. Shrinking the directory can be
implemented in the same way as doubling its size is
done. Merging buckets is more complex. The basic idea is the same as
for splitting: no update operations should be allowed to execute on buckets that
can be merged (in order to avoid violating linearizability). However,
there are
two main differences between merging and splitting: (i) merging is applied
to non-full buckets, and (ii) it involves more than one bucket.

To address the first point we use a mechanism to freeze a bucket,
similarly to what is proposed in \cite{Liu2014}. No update operations
can be run on a frozen bucket even if it is not full. Instead, the
thread willing to run an update operation on a frozen bucket should
help running the merging action first. We have implemented freezing
using a flag that is stored in the bucket state. The flag is modified
by a thread using \texttt{ApplyWFOp()}. To address the second point,
we perform merging in two steps: first, a thread tries to freeze all
buckets involved in the merging, and then it calls \texttt{ResizeWF()}
to perform the merging action.

Note that since the merging of buckets is done in several steps, it may fail:
a thread may not manage to freeze all buckets involved in the merging
action. The first reason why a thread may fail to freeze a bucket is
that the bucket might be full. The second reason is that the bucket might be already
frozen, e.g., because it is involved in
another merging action. This might happen if two threads want to
execute conflicting merging actions, for instance, one wants to create
a bucket with prefix $001$ and the other with prefix $0010$. Since the
two new buckets have a common prefix, the two merging actions involve
overlapping sets of buckets. One of the two threads will not manage to
freeze all required buckets for its action to take place. We remark that to
avoid having both conflicting merge actions fail, all threads should
freeze buckets in the same order. If a merge action fails, some buckets
might have to be unfrozen. This operation is done during a directory
update through the \texttt{ResizeWF()} function.
It is determined by the user when merging will be triggered.

\subsection{Correctness and Progress}

When no resizing occurs, linearizability 
is proved following similar arguments as those used to prove \PSIM\ correct.  
When resizing occurs, we prove that: 
(1)   each operation is executed exactly once, and (2)
  once an update operation has been executed, its 
  modifications cannot be lost due to resizing.

\LOOKUP\ operations are obviously wait-free. Each \INSERT\ operation
calls at most two functions (\texttt{ApplyWFOp()} and
\texttt{ResizeWF()}) that implement \PSIM. Since \PSIM\ is wait-free,
it remains to show that functions called by these two functions
execute a bounded number of instructions. 
\PSIM\ creates a copy of the state to manipulate. Hence, the
code executed by any instance of \PSIM\ is sequential. Note that
functions \texttt{ExecOnBucket()}, \texttt{SplitBucket()} and
\texttt{DirectoryUpdate()} execute a bounded number of
instructions. \texttt{ApplyWFOp()} calls \texttt{ExecOnBucket()} at
most $n$ times.
So,  the total number of executed
instructions is bounded in this case.

The loops at lines \ref{line:iterhelp} and \ref{line:iterhelpapp}
imply that the upper bound on the number of calls to
\texttt{ApplyPendingResize()} by \texttt{ResizeWF()} is 
$O(n^{2})$. The number of instructions run by
\texttt{ApplyPendingResize()} is bounded by the number of times a
bucket should be split in order to apply its pending operations. We
conclude that the total number of instructions executed is bounded.

\section{Implementation \label{sec:implem}}

The description provided in Section~\ref{sec:algo} assumes that a
garbage collector (GC) is available. 
As pointed out in previous studies~\cite{David2015, Gramoli2015},
memory management can have a severe impact on the performance of concurrent
data structures. So, we paid special attention to this issue
while we were implementing our algorithm. 
Specifically, we use an epoch-based
non-blocking garbage collector~\cite{Fraser2004}. It is
based on thread-local counters that are incremented when a thread
performs an operation. The value of these counters are checked
periodically to decide when some released memory can safely be
reused. The frequency at which counters are checked depends on the
size of the thread-local batches that are used to store references to
elements that have been released: when a batch becomes full, the counters are checked to
see if previous batches can be released.

Our implementation has to be able to efficiently allocate bucket
states (\texttt{BState}) since any update operation requires the
allocation of a new \texttt{BState} record. We implemented
thread-local memory pools (heaps) that are optimized to allocate
memory blocks of the size of a \texttt{BState} record. We also
implemented a mechanism to allow memory that has been released by the
GC to be inserted back in these thread-local heaps.

To further improve the performance of our algorithm in practice, we
apply an optimization to reduce the size of \texttt{BState}
objects. Namely, we reduce the size of the $results$ array 
by storing the actual results of the
operations in a separate shared array and storing in the
\texttt{BState} an array of integers corresponding to indices in this
shared array. To obtain the result of its executing operation, a
thread $i$ should read $results[i]$ in the \texttt{BState}, and then
access the appropriate element of the shared array. This shared array
is divided to non-overlapping blocks, one for each thread, so that 
each thread writes to its own block (with no need for synchronization).

We implemented the toggle vectors so that bits can be flipped
efficiently using atomic \ADD\ (see~\cite{Fatourou2013}).
Note also that when the \CAS\ of lines~\ref{line:CAS} is successful,
\texttt{ApplyWFOp()} can return immediately.

\section{Performance Evaluation\label{sec:eval}}

\subsection{Evaluated algorithms \label{sec:evaluatedAlg}}

We implemented our algorithm and the memory management component in
C. The implementation allows activating or deactivating the use of
local heaps for memory allocation. Our algorithm is called
\WFExt\ hereafter.

We compare the performance of \WFExt\ with the performance of the
lock-free implementations presented by Shalev and
Shavit~\cite{shalev2006} (called \LFSplit) and by Liu et
  al.~\cite{Liu2014} (called \LFFreeze). \LFSplit\ is the reference
implementation of the algorithm by Shalev and Shavit available
online\footnote{\url{http://www.memoryhole.net/kyle/2011/06/02/}}. \LFFreeze\ is
the C version of the code provided by the
authors\footnote{\url{https://github.com/mfs409/nonblocking/tree/master/tsx_acceleration/chash}}.
To ensure a fair comparison, we also created modified versions of
these algorithms. A modified version is identified with the suffix
\texttt{-U}. Specifically, for \LFSplit, we removed the global
counter, used for resizing decisions, that is accessed at the end of
the execution of each update operation. Instead, resizing decisions
are taken based on the size of individual buckets. For \LFFreeze, the
original version of the authors only stored keys. We modified it to
store key-value pairs. Also, in the original version, inserting a key
that is already present implies no modification of the hash table. In
our modified version, the value should be updated to comply with the
semantics of the operations we described in Section
\ref{sec:algo}. Note that this has a big impact on the performance
results because with this semantic no \INSERT\ boils down to a
\LOOKUP\ (This explains the differences between our graphs and the
ones presented in \cite{Liu2014}). Starting from these modified
versions, we also implemented a version of each algorithm that uses
our memory management component (marked with suffix {\sf -M}).
\LFFreezeM\ uses fixed-size buckets (instead of varying size buckets)
to better leverage our local heaps. In its original
  implementation, \LFSplit\ relies on the system memory allocator,
  while \LFFreeze\ relies on its own epoch-based GC and does not use local
  heaps.

Our evaluation also includes a simple \emph{blocking} 
  non-resizable hash table (called \Lock)
  where each bucket is protected using a lock and each operation
 on the bucket has to acquire this lock.

\subsection{Experimental setup}

We have performed experiments in two multiprocessor architectures: a $48$-core
machine equipped with $4$ Intel Haswell-EX E7-4830-v3 processors and a
$64$-core machine equipped with $4$ Intel Broadwell E7-4850-v4
processors. All experiments showed similar performance behavior in both
machines. Hence, we only present the results of our experiments on the larger
and newer $64$-core machine. This machine features 256GB of RAM
distributed over 4 NUMA nodes. The operating system is Linux with kernel
4.9.9 and the compiler is \texttt{gcc} 5.4. All codes are compiled at
the maximum level of optimization. All tests were run 8 times with a 5-second
duration per test. We present the average throughput over all runs.

As discussed in other studies, the underlying memory allocator has a
big performance impact. All evaluations were run with two allocators:
the \emph{glibc} allocator and the TCMalloc
allocator\footnote{\url{http://goog-perftools.sourceforge.net/doc/tcmalloc.html}}.
For each experiment and each algorithm, we display results for the
allocator that provides the best throughput. Information about the
selected allocator is provided in the legend of the graphs:
"\texttt{G}" for the \emph{glibc} allocator and "\texttt{T}" for
TCMalloc. We also test the two main NUMA policies for physical memory
allocation, that is \emph{local} (named "\texttt{L}" in graphs'
legends) and \emph{interleave} (named "\texttt{I}"), and display the
best result. The NUMA policy is enforced using the \emph{numactl}
program.

Several parameters can be configured for each algorithm, namely, the
bucket size, the size of the batches used to store released memory
records before running the Garbage Collector (GC), and the size of the
thread-local heaps. We performed a wide range of experiments to
identify parameters that suits all algorithms. We use these parameters
throughout all experiments: buckets of size 8, GC batches of size 256,
thread-local heaps of size 8.

Finally, in all experiments, the operations to be executed, as well as
their parameter (the key), are randomly selected using the
TinyMT\footnote{https://github.com/MersenneTwister-Lab/TinyMT}
pseudo-random number generator. This generator produces numbers with a
uniform distribution.

\subsection{Directory-stable performance}

First, we study the performance of hash tables in the most
common scenario, that is, when resizing is rare.  More specifically,
we evaluate the throughput of the implementations starting from a hash
table where we have already inserted key-value pairs corresponding to
half of the keys manipulated during the experiment and we run
workloads with the same amount of \DELETE\ and \INSERT\ operations.

\begin{figure}[t]
  \begin{center}
    \subfloat[50\% \LOOKUP\ operations.\label{fig:steady_half5}]{
      \includegraphics[width=0.50\hsize]{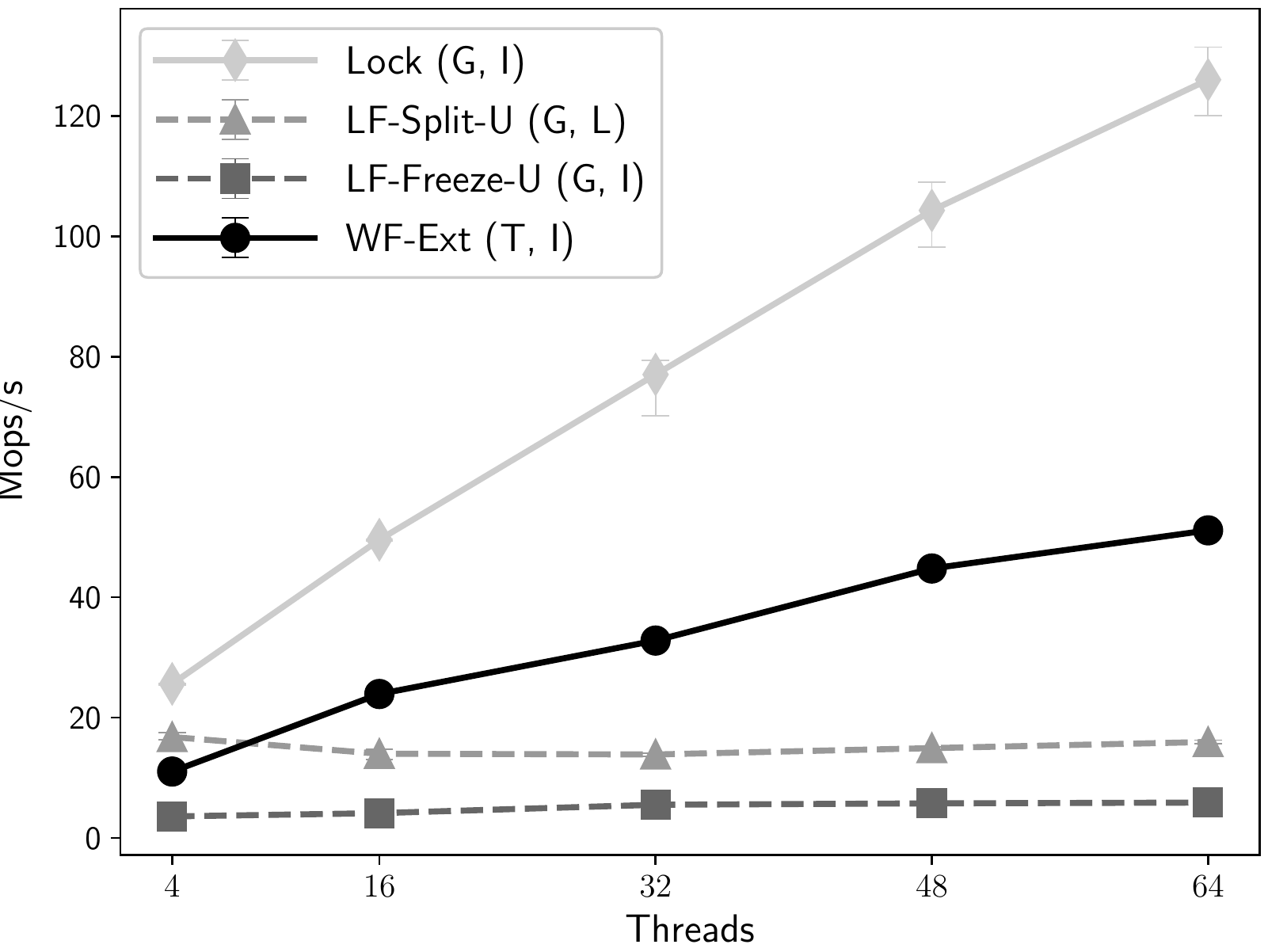}
    }
    \subfloat[90\% \LOOKUP\ operations.\label{fig:steady_half9}]{
      \includegraphics[width=0.50\hsize]{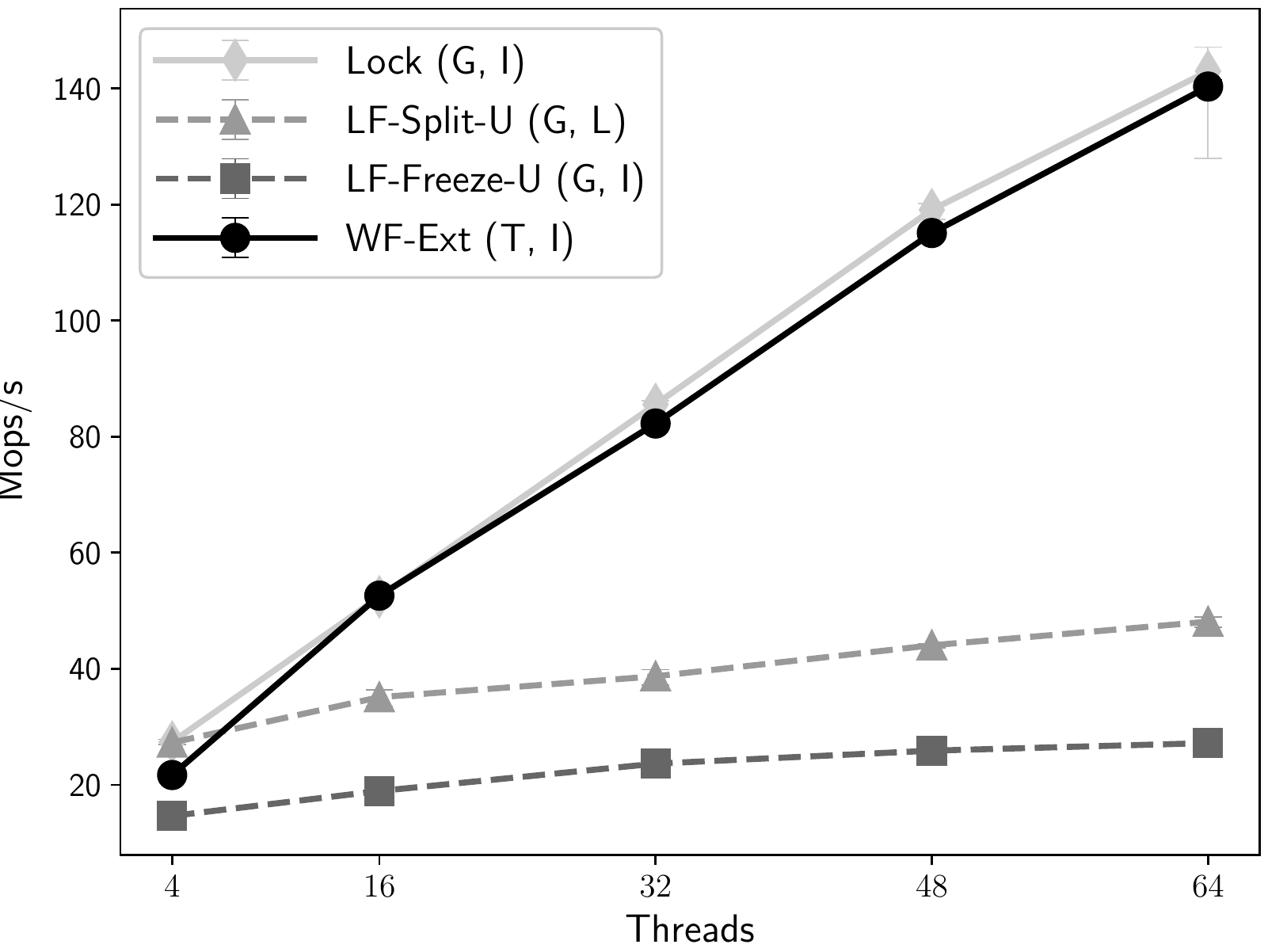}
    }
    \caption{Directory-stable state throughput in the Intel 64-core
      machine using $1K$ items. No local
      heaps. \label{fig:steady_half}}
  \end{center}
\end{figure}

Our first test uses $1K$ keys. Figures \ref{fig:steady_half5} and
\ref{fig:steady_half9} correspond to loads with 50\% and 90\% of
\LOOKUP\ operations, respectively. The rest of the
operations are equally divided between \INSERT\ and \DELETE. The
figures compare the performance of our wait-free algorithm with the
performance of the two modified lock-free algorithms \LFSplitU\ and
\LFFreezeU, when relying on the underlying system memory allocator
(\emph{glibc} malloc or TCMalloc) for memory allocation. We choose to
display \LFSplitU\ because it constantly outperforms the original
\LFSplit\ algorithm (which confirms the importance of rule (B)). We
choose to display \LFFreezeU\ because a comparison with
\LFFreeze\ would not be fair due to the differences
in the implementation of \INSERT\ (see Section \ref{sec:evaluatedAlg}).

In Figure~\ref{fig:steady_half}, our hash table outperforms
the two lock-free algorithms. The high throughput of our algorithm is
mostly due to highly efficient \LOOKUP\. This is
confirmed by the large improvement observed when
increasing the percentage of \LOOKUP s to 90\%. We attribute the lower
performance of the two lock-free algorithms to different
factors. Specifically, for \LFSplitU, the reason is twofold: (i) the
algorithm does not comply with design rule (A), implying less
efficient \LOOKUP\ operations in this case; (ii) the buckets are
implemented as lists of items, which makes item search less efficient
than with buckets implemented as arrays. In the case of \LFFreezeU\,
the main reason we identified for the lower performance is that the
bucket size is not fixed. It is adapted to the number of items to
store. Since a new bucket should be allocated for each update
operation, it creates an unfriendly workload for the memory allocation
system.

\begin{figure}[t]
  \begin{center}
    \subfloat[50\% \LOOKUP\ operations.\label{fig:steady_pool_half5}]{
      \includegraphics[width=0.50\hsize]{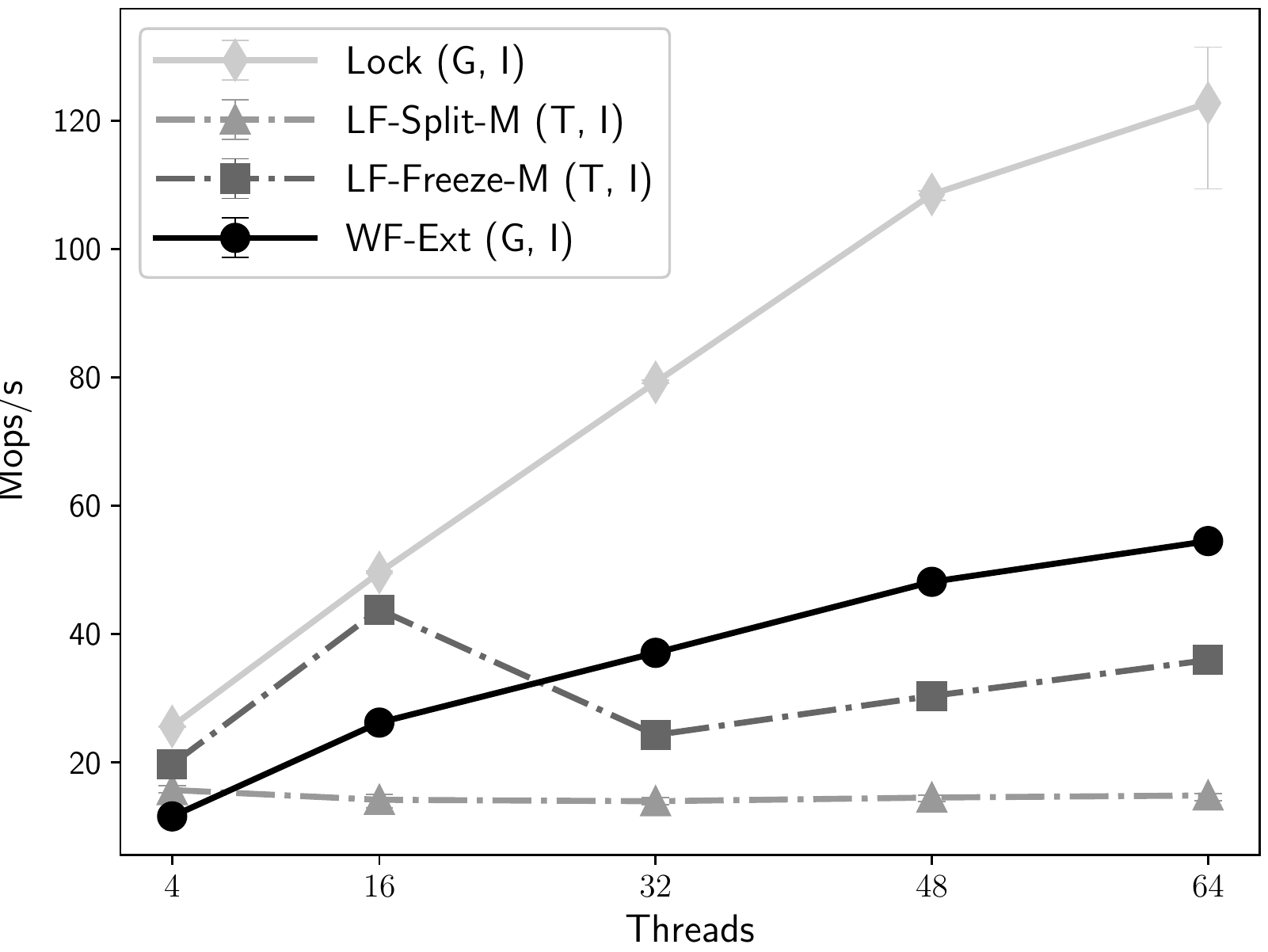}
    }
    \subfloat[90\% \LOOKUP\ operations.\label{fig:steady_pool_half9}]{
      \includegraphics[width=0.50\hsize]{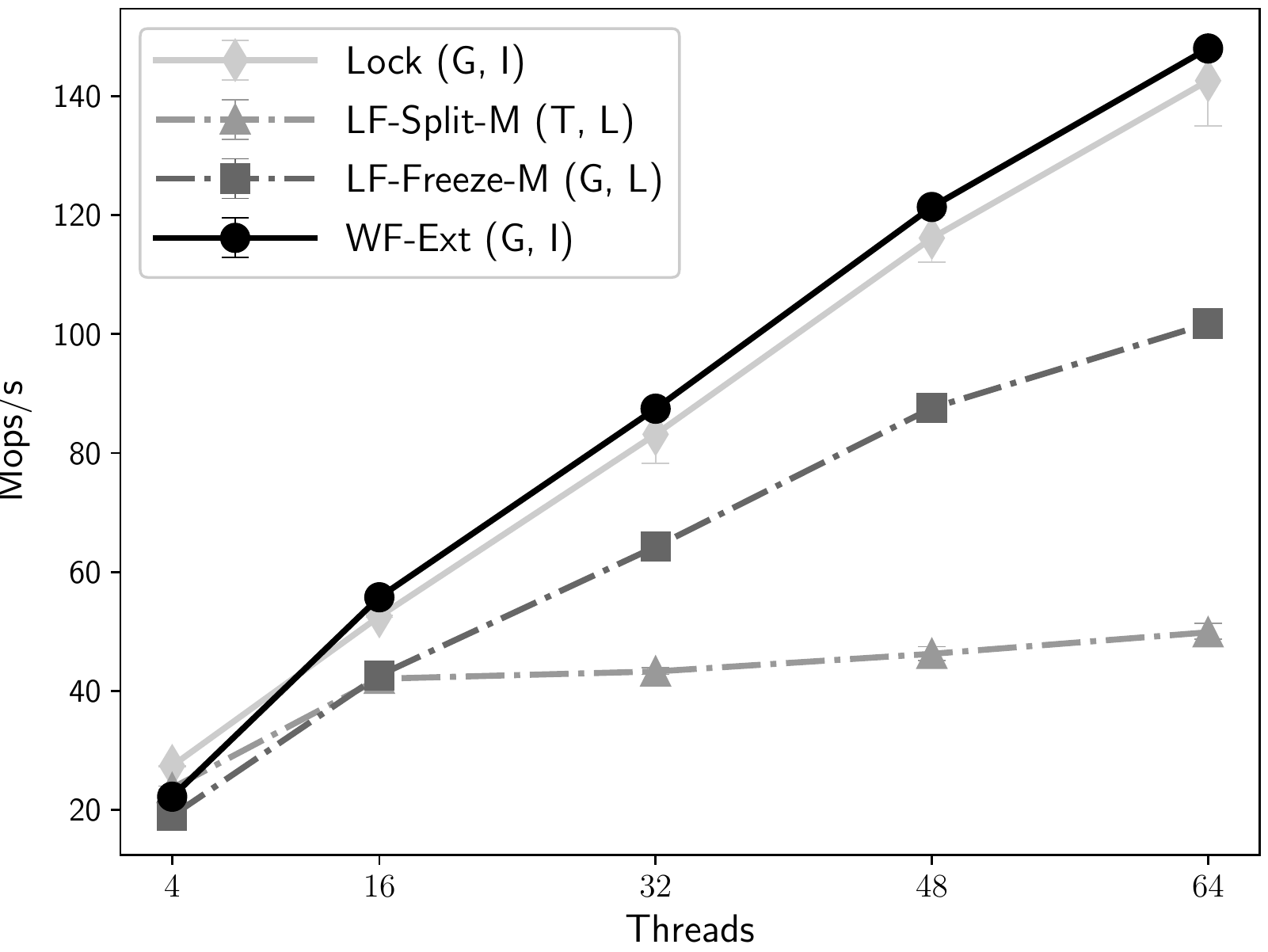}
    }
    \caption{Directory-stable state throughput in the Intel 64-core
      machine, using $1K$ keys. With local
      heaps. \label{fig:steady_pool_half}}
  \end{center}
\end{figure}

Figure~\ref{fig:steady_pool_half} presents the same experiment as in
Figure~\ref{fig:steady_half}, this time using both our GC and the
local heaps for all non-blocking algorithms ({\sf -M} suffix). This
figure shows that even when compared using the exact same memory
management component, \WFExt\ is more efficient than the lock-free
algorithms. In this experiment, with 90\% \LOOKUP s, our algorithm is
up to 47\% better than the second best non-blocking algorithm. Still,
the performance of \LFFreezeM\ shows that using our memory management
component and fixed-size buckets greatly improves the performance of
\LFFreeze. On the other hand, \LFSplit\ does not benefit from using
local heaps.

\begin{figure}[t]
  \vspace{-0.3cm}
  \begin{center}
    \subfloat[50\% \LOOKUP\ operations.\label{fig:steady_large_pool_half5}]{
      \includegraphics[width=0.50\hsize]{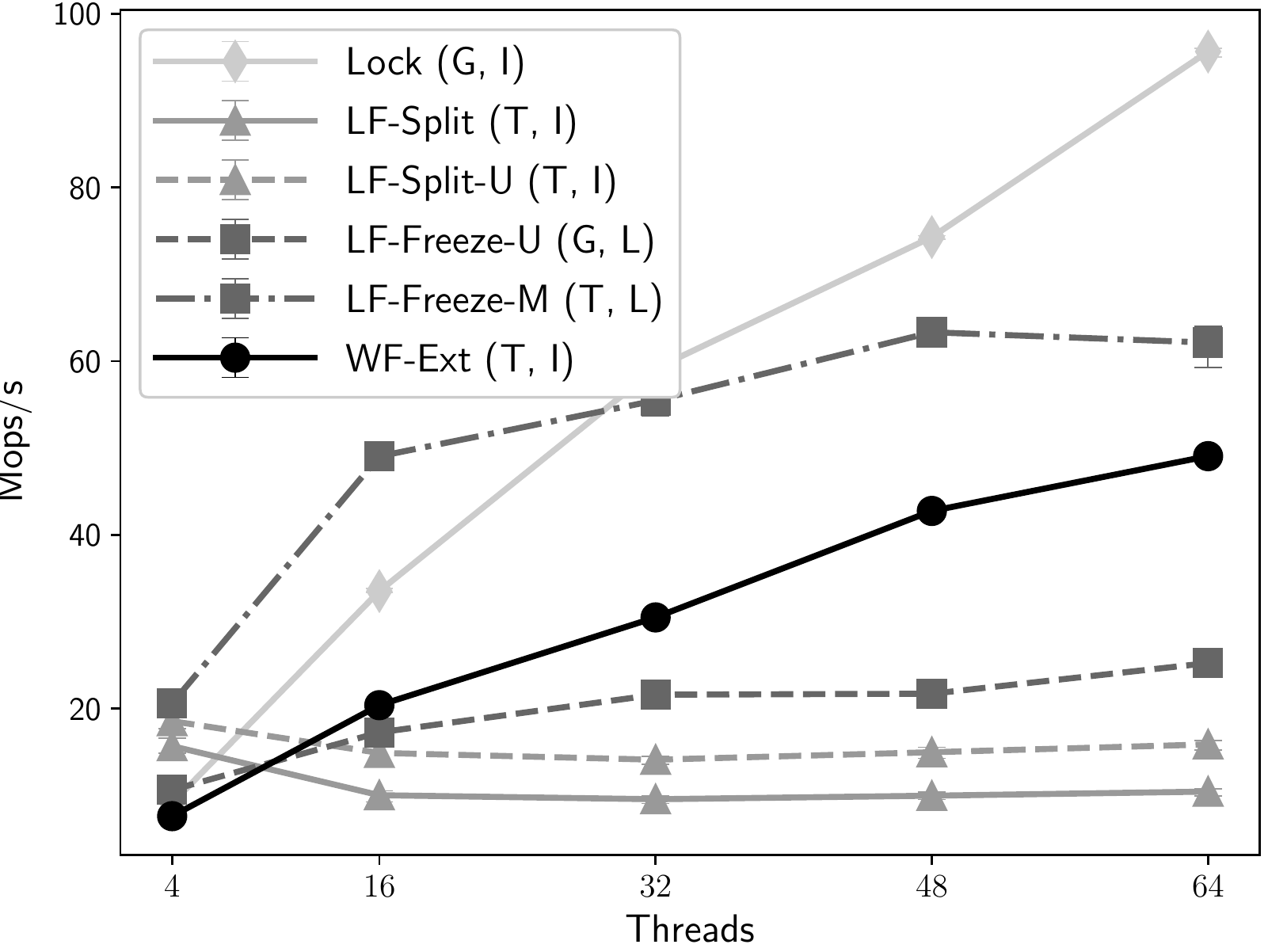}
    }
    \subfloat[90\% \LOOKUP\ operations.\label{fig:steady_large_pool_half9}]{
      \includegraphics[width=0.50\hsize]{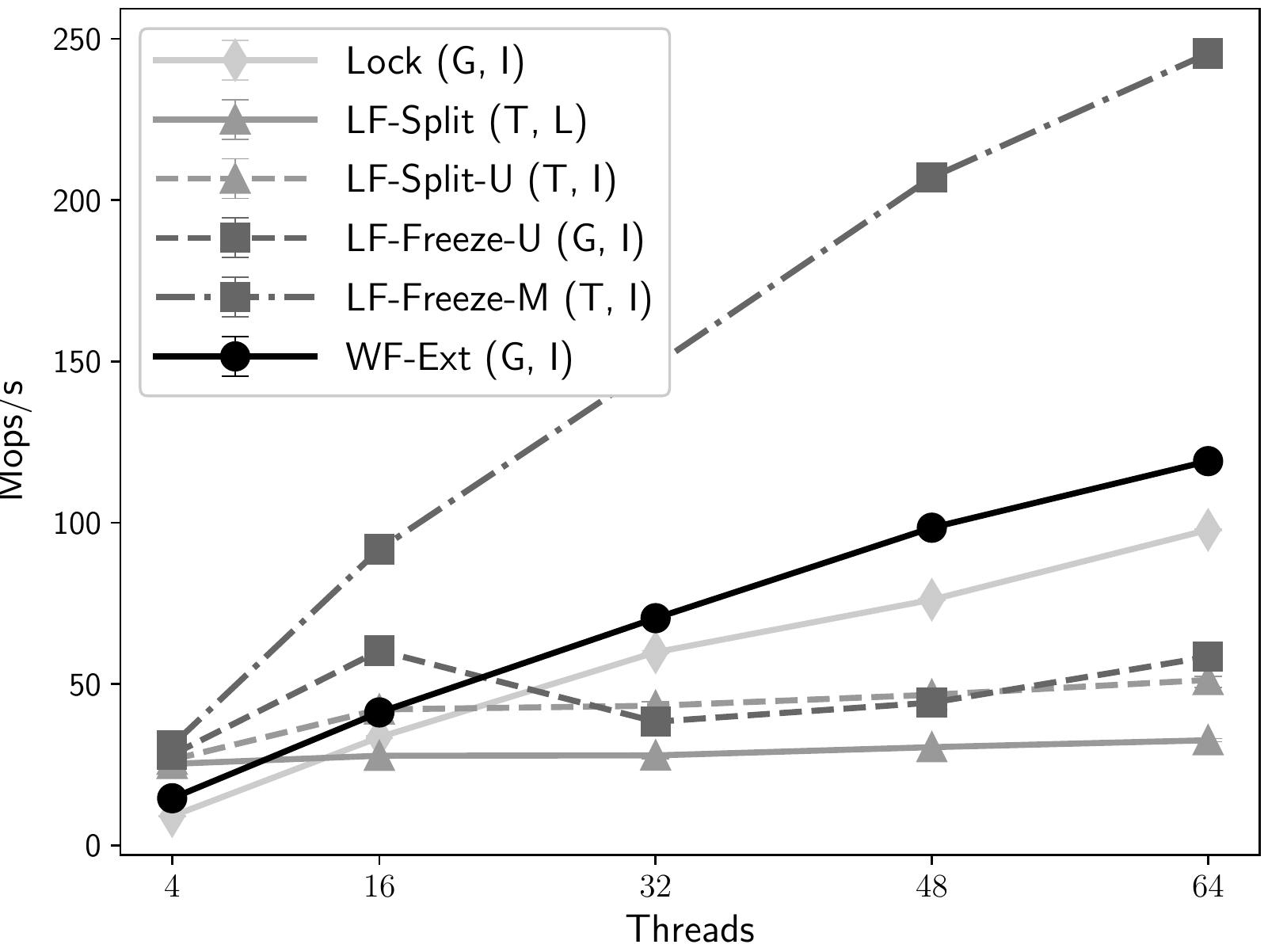}
    }
    \caption{Directory-stable state throughput in the Intel 64-core
      machine, using $256K$ keys. \label{fig:steady_large_pool_half}}
  \end{center}
  \vspace{-0.3cm}
\end{figure}

To assess the performance of our algorithm when storing a larger
number of items, Figure~\ref{fig:steady_large_pool_half} presents the
same experiment as in Figure~\ref{fig:steady_pool_half}, this time
when keys are selected among $256K$ keys. For the sake of
completeness, we present the performance of two versions of each
lock-free algorithm. For \LFSplit, we present the original version and
\LFSplitU\ (\LFSplitM\ is not considered because it does not reach the
performance of \LFSplitU). For the other algorithm, we present both
\LFFreezeU\ and \LFFreezeM.

Compared to the algorithms described in the related work (original
algorithms or \texttt{-U} versions), the performance of our algorithm
is again by far the best. It even outperforms the lock-based
hash-table when the percentage of \LOOKUP s is high. The high
performance of \LFFreezeM\ demonstrates that the algorithm proposed by
Liu et al. can benefit from the modifications we suggest to
efficiently manage memory allocation. Once memory management issues
are solved, \LFFreezeM\ outperforms \WFExt\ because its update
operations are less complex since it implements a weaker progress
condition.

The fact that \WFExt\ outperforms all lock-free implementations when
the hash table has a small number of buckets can be surprising,
especially since existing wait-free resizable hash tables are
performing much worse than their lock-free
counterparts~\cite{Liu2014}. This result is due to the use of the
\PSIM\ universal construction~\cite{Fatourou2013}. Indeed,
\PSIM\ performs especially well on contented objects since it employs
the technique of \emph{combining} the operations of different threads
that are applied to the same object~\cite{Hendler2010}.

\subsection{Resizing efficiency}

\begin{figure}[t]
  \begin{center}
    \subfloat[Resizing efficiency\label{fig:resizingtime}]{
      \includegraphics[width=0.5\hsize]{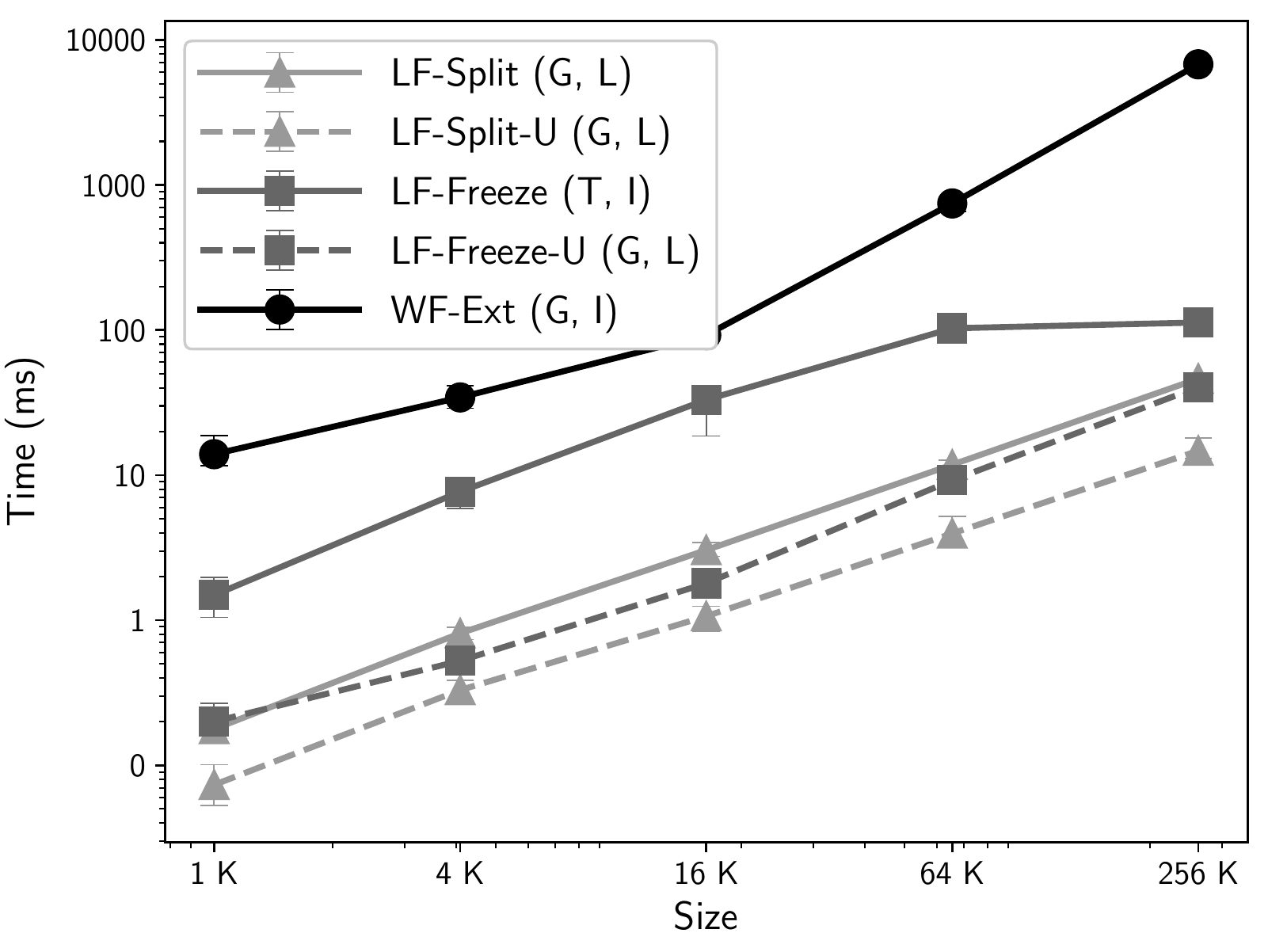} }
    \subfloat[Throughput of 5-second runs with 1k
      items\label{fig:resizingamortive}]{
      \includegraphics[width=0.5\hsize]{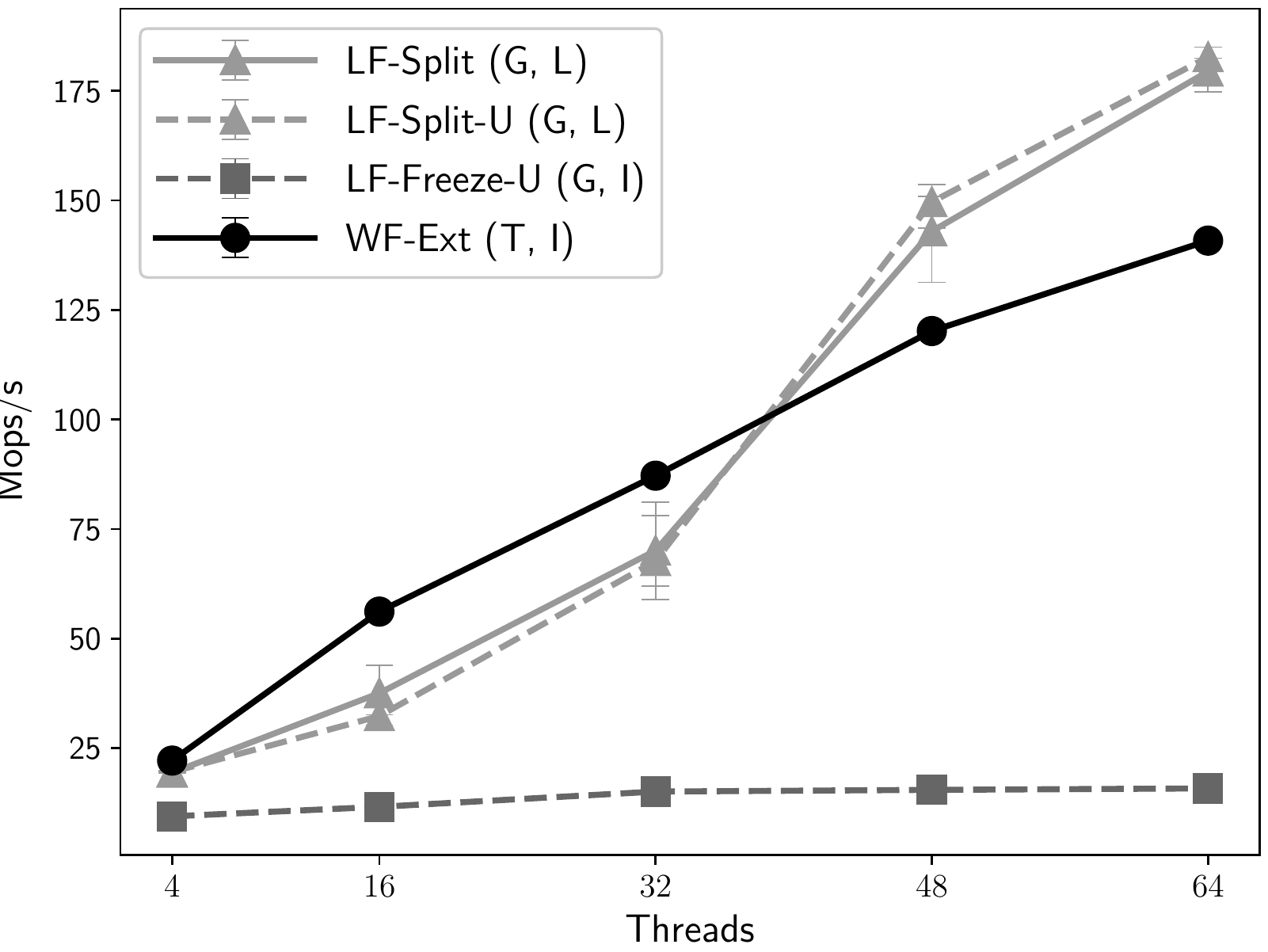}
    }
    \caption{Performance with Resizing. Tests with 64 threads\label{fig:resizing}}
  \end{center}

\end{figure}

To evaluate resizing performance, we run a test where the hash table
starts with only $2$ buckets. Multiple threads start inserting items
randomly, and we measure the time it takes for the hash table to reach
its final size. To have a more realistic workload, threads also
execute \LOOKUP\ operations on the items with a 50\%
probability. Figure~\ref{fig:resizingtime} presents the results. Note
that to make the figure readable, we use log scales on both axis.

The figure shows that the resizing performance of our algorithm is
much lower than that of evaluated lock-free algorithms. However, the
experiment of Figure \ref{fig:resizingamortive} illustrates that
\WFExt\ still performs well under resizing, if the cost of resizing is
amortized over long runs. In this experiment, we evaluate the
throughput on a $5$-second run when starting from a hash table with
$2$ buckets and manipulating $1K$ items in a load with 90\%
\LOOKUP\ and 10\% \INSERT\ operations. In this case, \WFExt\ reaches
the same throughput as when resizing actions are rare
(Figure~\ref{fig:steady_pool_half9}). Thus, in sufficiently long runs
the performance impact of resizing is not significant. We note that
\LFSplit\ achieves much better performance than in previous
experiments. Indeed, in this test, \LOOKUP s for \LFSplit\ are more
efficient since no item is ever deleted, confirming the validity of
design rule (A).

\section{Conclusion}

This paper presents a resizable wait-free hash table based on
extendible hashing. The design of our algorithm is based on two design
rules that aim at preserving the natural parallelism of concurrent
hashing in the most common case, that is, when no resizing action
occurs. Leveraging several instances of the \PSIM\ universal
construction to ensure wait-freedom, our implementation achieves
unprecedented performance for a wait-free resizable hash table. More
generally, our approach provides a new trade-off in the performance of
resizable hash tables. Namely, experiments run on large multicore
architectures, show that, at the cost of more expensive resizing
actions, our algorithm largely outperforms lock-free hash tables
described in related studies when resizing actions are rare, while
providing a stronger progress guarantee.

\section*{Acknowledgments}
Part of this work has received funding from the European Union's
Horizon 2020 research and innovation programme under grant agreement
No 671578. The authors would like to thank Darko Petrovi\'{c} and
Andr\'{e} Schiper for following the initial stages of this work and
providing feedback.

\bibliographystyle{abbrv}
\bibliography{ms}

\begin{thebibliography}{10}

\bibitem{Boyd-Wickizer2010}
S.~Boyd-Wickizer, A.~T. Clements, Y.~Mao, A.~Pesterev, M.~F. Kaashoek,
  R.~Morris, and N.~Zeldovich.
\newblock An analysis of linux scalability to many cores.
\newblock In {\em Proceedings of the 9th USENIX Conference on Operating Systems
  Design and Implementation}, OSDI'10, pages 1--16, 2010.

\bibitem{David2015}
T.~David, R.~Guerraoui, and V.~Trigonakis.
\newblock Asynchronized concurrency: The secret to scaling concurrent search
  data structures.
\newblock In {\em Proceedings of the Twentieth International Conference on
  Architectural Support for Programming Languages and Operating Systems},
  ASPLOS '15, pages 631--644, 2015.

\bibitem{Ellis1983}
C.~S. Ellis.
\newblock Extendible hashing for concurrent operations and distributed data.
\newblock In {\em Proceedings of the 2Nd ACM SIGACT-SIGMOD Symposium on
  Principles of Database Systems}, pages 106--116, 1983.

\bibitem{Enbody1988}
R.~J. Enbody and H.~C. Du.
\newblock Dynamic hashing schemes.
\newblock {\em ACM Computing Surveys}, 20(2), July 1988.

\bibitem{Fagin1979}
R.~Fagin, J.~Nievergelt, N.~Pippenger, and H.~R. Strong.
\newblock Extendible hashing -- a fast access method for dynamic files.
\newblock {\em ACM Transactions on Database Systems}, 4(3):315--344, Sept.
  1979.

\bibitem{Fatourou2012}
P.~Fatourou and N.~D. Kallimanis.
\newblock Revisiting the combining synchronization technique.
\newblock In {\em Proceedings of the 17th ACM SIGPLAN symposium on Principles
  and Practice of Parallel Programming}, 2012.

\bibitem{Fatourou2013}
P.~Fatourou and N.~D. Kallimanis.
\newblock Highly-efficient wait-free synchronization.
\newblock {\em Theory of Computing Systems}, pages 1--46, 2013.

\bibitem{Feldman2013}
S.~Feldman, P.~LaBorde, and D.~Dechev.
\newblock Concurrent multi-level arrays: Wait-free extensible hash maps.
\newblock In {\em International Conference on Embedded Computer Systems:
  Architectures, Modeling, and Simulation}, pages 155--163, 2013.

\bibitem{Fraser2004}
K.~Fraser.
\newblock {\em Practical Lock-Freedom}.
\newblock PhD thesis, University of Cambridge, 2004.

\bibitem{Gramoli2015}
V.~Gramoli.
\newblock More than you ever wanted to know about synchronization:
  Synchrobench, measuring the impact of the synchronization on concurrent
  algorithms.
\newblock In {\em Proceedings of the 20th ACM SIGPLAN Symposium on Principles
  and Practice of Parallel Programming}, PPoPP 2015, 2015.

\bibitem{Greenwald2002}
M.~Greenwald.
\newblock Two-handed emulation: How to build non-blocking implementations of
  complex data-structures using dcas.
\newblock In {\em Proceedings of the 21st Annual Symposium on Principles of
  Distributed Computing}, PODC '02, 2002.

\bibitem{Hendler2010}
D.~Hendler, I.~Incze, N.~Shavit, and M.~Tzafrir.
\newblock Flat combining and the synchronization-parallelism tradeoff.
\newblock In {\em Proceedings of the 22nd ACM symposium on Parallelism in
  algorithms and architectures}, 2010.

\bibitem{Herlihy1991}
M.~Herlihy.
\newblock Wait-free synchronization.
\newblock {\em ACM Transactions on Programming Languages and Systems},
  13(1):124--149, Jan. 1991.

\bibitem{Herlihy2008}
M.~Herlihy and N.~Shavit.
\newblock {\em {The Art of Multiprocessor Programming}}.
\newblock Morgan Kaufmann Publishers Inc., 2008.

\bibitem{Herlihy1990}
M.~Herlihy and J.~Wing.
\newblock Linearizability: a correctness condition for concurrent objects.
\newblock {\em ACM Transactions on Programming Languages and Systems},
  12(3):463--492, July 1990.

\bibitem{Jenkins2017}
L.~Jenkins, T.~Zhou, and M.~Spear.
\newblock {Redesigning Go's Built-In Map to Support Concurrent Operations}.
\newblock In {\em 26th International Conference on Parallel Architectures and
  Compilation Techniques (PACT)}, 2017.

\bibitem{Kogan2012}
A.~Kogan and E.~Petrank.
\newblock A methodology for creating fast wait-free data structures.
\newblock In {\em Proceedings of the 17th ACM SIGPLAN Symposium on Principles
  and Practice of Parallel Programming}, PPoPP '12, 2012.

\bibitem{Lea2003}
D.~Lea.
\newblock Package util.concurrent release 1.3.4 (2003).
\newblock
  http://gee.cs.oswego.edu/dl/classes/EDU/oswego/cs/dl/util/concurrent/intro.html.

\bibitem{Liu2014}
Y.~Liu, K.~Zhang, and M.~Spear.
\newblock {Dynamic-sized Nonblocking Hash Tables}.
\newblock In {\em Proceedings of the 2014 ACM Symposium on Principles of
  Distributed Computing}, PODC '14, 2014.

\bibitem{Michael2002}
M.~M. Michael.
\newblock High performance dynamic lock-free hash tables and list-based sets.
\newblock In {\em Proceedings of the 14th Annual ACM Symposium on Parallel
  Algorithms and Architectures}, SPAA '02, 2002.

\bibitem{shalev2006}
O.~Shalev and N.~Shavit.
\newblock Split-ordered lists: Lock-free extensible hash tables.
\newblock {\em Journal of the ACM}, 53(3):379--405, 2006.

\bibitem{Shavit2011}
N.~Shavit.
\newblock Data structures in the multicore age.
\newblock {\em Communications of the ACM}, 54(3):76--84, Mar. 2011.

\bibitem{Triplett2011}
J.~Triplett, P.~E. McKenney, and J.~Walpole.
\newblock Resizable, scalable, concurrent hash tables via relativistic
  programming.
\newblock In {\em USENIX Annual Technical Conference}, 2011.

\end{thebibliography}

\end{document}